\documentclass[final,fleqn,3p,twoside]{elsart}
\usepackage{graphicx}
\usepackage{dcolumn}
\usepackage{bm}
\usepackage{float}
\usepackage{amsfonts,amsmath,amssymb}
\usepackage{pst-grad,color}
\usepackage[bookmarks]{hyperref}

\newcommand{\up}{\uparrow}
\newcommand{\down}{\downarrow}
\def\O{\mathcal{O}}
\def\Or{\mathcal{O}}
\def\U{\mathcal{U}}

\def\C{\mathcal{C}}

\newcommand{\unity}{1 \hskip-2.5pt{\rm l}}
\def\pd#1#2{\frac{\partial#1}{\partial #2}} 
\newcommand{\be}{\begin{equation} } 
\newcommand{\ee}{\end{equation}   }
\def\normord#1{\ensuremath{\mathinner{%
    \mathopen{\boldsymbol:}#1\mathclose{\boldsymbol:}}}}
\def\commutator#1#2{\ensuremath{\mathinner{%
    \mathopen[#1,#2\mathclose]}}}
\def\anticommutator#1#2 {\ensuremath{\mathinner{%
      \left\lbrace #1,#2 \right\rbrace }}}%
\def\bra#1{\ensuremath{\left\langle {#1}\right\vert}}
\def\ket#1{\ensuremath{\left\vert{#1}\right\rangle }}
\def\bracket#1#2{\ensuremath{\left\langle{#1}\vert{#2}\right\rangle }}

\def\abs#1{\mathinner{\lvert#1\rvert}}

\journal{Annals of Physics}

\begin{document}

\begin{frontmatter}

\title{Real-time evolution for weak interaction quenches in quantum systems}

\author[ASC]{Michael Moeckel},
\author[ASC]{Stefan Kehrein}
\ead{Michael.Moeckel@physik.lmu.de}

\ead{Stefan.Kehrein@physik.lmu.de}

\address[ASC]{Arnold-Sommerfeld-Center for Theoretical Physics, Center for NanoSciences and Department f\"ur Physik, Ludwig-Maximilians-Universit\"at M\"unchen, Theresienstra\ss{}e 37, 80333 M\"unchen, Germany}


\begin{abstract} 
Motivated by recent experiments in ultracold atomic gases that explore the nonequilibrium dynamics of interacting quantum many-body systems, we investigate the nonequilibrium properties of a Fermi liquid. We apply an interaction quench within the Fermi liquid phase of the Hubbard model by switching on a weak interaction suddenly; then we follow the real-time dynamics of the momentum distribution  by a systematic expansion in the interaction strength based on the flow equation method  \cite{Moeckel2008}. In this paper we derive our main results, namely the applicability of a quasiparticle description, the observation of a new type of quasi-stationary nonequilibrium Fermi liquid like state and a delayed thermalization of the momentum distribution. 
We explain the physical origin of the delayed relaxation as a consequence of phase space constraints in fermionic many-body systems. This brings about a close relation to similar behavior of one-particle systems which we illustrate by a discussion of the squeezed oscillator; we generalize to an extended class of systems with discrete energy spectra and point out the generic character of the nonequilibrium Fermi liquid results for weak interaction quenches. Both for discrete and continuous systems we observe that particular nonequilibrium expectation values are twice as large as their corresponding analogues in equilibrium. For a Fermi liquid, this shows up as an increased correlation-induced reduction of the quasiparticle residue in nonequilibrium.
\end{abstract}
\end{frontmatter}

\section{Introduction}

\subsection{Experimental approaches}
Recent experimental progress has stimulated the investigation of interacting many-particle quantum systems in nonequilibrium. In 1998 ultracold atoms confined by a harmonic trapping potential have first been loaded \cite{Anderson1998} onto arrays of standing light waves known as optical lattices \cite{Jessen1996}. Since that time the implementation of  paradigmatic quantum models of condensed matter physics, foremost the Hubbard model \cite{Hubbard1963}, in such systems with long coherence times and high tunability became a feasible task. While in solids parameters like the lattice spacing or the interaction strength are fixed constants predefined by nature, optical lattices provide a technique to experimentally study consequences of their time-dependent variation \cite{Zwerger2007}. The observation of the equilibrium Mott-Hubbard phase transition in the 3d bosonic Hubbard model \cite{Greiner2002_1} was soon followed by the discovery of  \emph{collapse and revival} phenomena when the system was quenched between its phases, i.e. when sudden changes were applied to the particle interaction \cite{Greiner2002_2}. Similar nonequilibrium behavior was found for quenched 1d hard-core bosons \cite{Kinoshita2006} which continued to oscillate without relaxation. 

Experiments with ultracold fermions had to overcome more technical difficulties. 
The observation of a Fermi surface in optical lattices \cite{Koehl2004}, of fermionic correlations \cite{Rom2006}, of superfluidity \cite{Chin2006} and the study of interaction-controlled transport by a quench in the trapping potential \cite{Strohmaier2007} prepared the recent observation of the Mott-Hubbard transition in the repulsive fermionic Hubbard model \cite{Joerdens2008}. This rapid progress of experimental sophistication gives hope that the predictions presented here will be subject to experimental observation in the near future.

\subsection{Thermalization debate}
Such new experimental opportunities and pioneering results have provoked further theoretical investigations into the nonequilibrium dynamics of well-established many-body model systems, e.g. for spin models  \cite{Barouch1970, Sengupta2003, Zurek2005, Calabrese2005, Cherng2006, Barmettler2008}, BCS systems  \cite{Yuzbashyan2005,Yuzbashyan2006b,Yuzbashyan2006, Warner2005}, 
1D hard-core bosons  \cite{Gangardt2007, Rigol2006A, Rigol2007}, a Luttinger liquid  \cite{Cazalilla2006, Cazalilla2009}, the Richardson model  \cite{Faribault2008}, and the Falicov-Kimball model  \cite{Eckstein2008}.
Excited by a quantum quench, i.e. by a sudden switch-on of the interaction term in the Hamiltonian, each of them exhibits its individual dynamics; however, they share the common feature that in many cases thermalization, i.e. the relaxation of time-averaged quantities towards a thermal ground state, has not been found. This observation has been linked to the integrability of these models which restricts a full relaxation due to additional conserved integrals of motion. Nonetheless, numerical studies report that breaking integrability of strongly correlated fermions does not lead to relaxation \cite{Manmana2007, Manmana2008}, and that the 1D Bose-Hubbard model, which is commonly assumed to be nonintegrable, does not equilibrate for certain sets of initial conditions \cite{Kollath2007, Kollath2008}.
 
This revitalized an ongoing debate about the long-term equilibration behavior of many-body quantum systems. It mimicked a similar discussion in nonlinear classical mechanics \cite{Ford1992} which followed a seminal paper by Fermi, Pasta and Ulam \cite{FermiCollectedPapers}. There the question was raised whether an excited system of coupled anharmonic oscillators would finally \emph{thermalize}, i.e. approach a steady state thermal distribution of the initial excitation energy onto all oscillator modes. To the surprise of the original authors in 1955 this has not been confirmed by their numerical calculations. 

In the \emph{classical case}, thermalization is regarded as a consequence of nonlinearities in the equations of motion which generate chaotic behavior. Then, in general, the orbit of the classical motion samples completely the hypersurface of the phase space that corresponds to all configurations which respect the conservation of certain extensive conserved quantities; then the system is called ergodic\footnote{A  classical system is called \emph{ergodic} if two averages of the phase space coordinates coincide:  a) The statistical (thermal) average over an ensemble of phase space configurations constrained to a hypersurface in phase space by constants of the motion (e.g. energy); and b) the time average of their dynamical solutions subjected to the same constants of the motion. 
This coincidence of a statistical (i.e. probabilistic) description of a physical system and a dynamical one (governed by deterministic laws of motion) is \emph{assumed} by Boltzmann's hypothesis.  Nonergodic cases can be easily found, e.g. systems with closed orbit solutions. It is an important result of nonlinear classical mechanics that nonlinearity in the equations of motions alone does not imply ergodicity.} and thermalization is expected. 
Although the FPU problem is still debated \cite{Flach2005}, the predominant opinion today is that Arnol'd diffusion will finally lead to thermalization of the system.

The time evolution of a \emph{quantum system} is governed by the linear Schr\"odinger equation; it can always be represented by a strictly unitary operation $\mathcal{U}(t,t_0)$ which is generated by the Hamiltonian. The later acts on the Hilbert space, such that an initial state is mapped onto a time-parametrized orbit of states constrained by constants of the motion like particle number, energy, etc. The time evolution of states is reflected in the evolution of expectation values of observables;  those correspond to measurable quantities. Note, for instance, that although the long time average of the dynamics of any quantum state equals its projection onto the zero energy (ground) state(s)\footnote{This can be easily seen by applying the time evolution operator in an eigenstate representation of the time-evolved state.}, this does not indicate a long-time relaxation towards (one of) its equilibrium ground state(s).

Moreover, the strict linearity of quantum mechanical time evolution implies a fundamental disagreement between a quantum mechanical description based on the Schr\"odinger equation and a quantum statistical one by means of a statistical operator $\hat{\rho}$; it is rooted in the additional coherence properties of a quantum system. 
Note that we only consider strictly closed quantum systems for which no tracing out of degrees of freedom related to an environment takes place.  
While a thermal state is constructed as an incoherent mixture of weighted quantum states with $\text{Tr}[\hat{\rho}_M^2] < 1$, a quantum system, once initialized in a pure state defined by $1\equiv\text{Tr}[\hat{\rho}_P^2] =\text{Tr}[\mathcal{U^{\dagger}}(t,t_0)\hat{\rho}_P^2\mathcal{U}(t,t_0)]$ remains pure at any later point in time because of the cyclic property of the trace\footnote{Note that the cyclic argument is not applicable for expectation values of observables $\langle \O \rangle = \text{tr}[\mathcal{U^{\dagger}}(t,t_0)\hat{\rho}_P\mathcal{U}(t,t_0) \O]$.}. Due to the linearity of the trace, the same holds for any time average. Hence quantum systems are never ergodic in a classical sense.  

This limitation does not apply for the expectation value of a particular observable. Nonetheless, the equilibration of expectation values in an integrable system is restricted for a different reason: Additional conserved integrals of motion arise from an exact integration of the equations of motion which prevent a wipe-out of the influence of the initial conditions such that thermalization with respect to a conventional Gibbs ensemble should not be expected. In analogy to equilibrium approaches \cite{Jaynes1957}, it was suggested that the long-time behavior can be reproduced by a statistical description based on a generalized Gibbs ensemble \cite{Rigol2007}. Many of the mentioned results \cite{Cazalilla2006, Rigol2007, Eckstein2008} explicitly agree with this approach and its prerequisites and limitations have been discussed \cite{Barthel2008, Gangardt2007}. A different notion of local relaxation grounds the examinations of finite subsystems \cite{Cramer2008a, Cramer2008b} which may exhibit thermal signatures even if the full system has not relaxed.

Yet for closed nonintegrable many-body systems the fundamental questions became more obvious: Which observables exhibit thermal long-time behavior, and for what reason? How does the large number of degrees of freedom present in a many-body system make up for the unitarity of time evolution such that a thermal long-time limit can appear?  
Some earlier works which addressed these questions by introducing an eigenstate thermalization hypothesis \cite{Deutsch1991, Srednicki1994} found new attention recently \cite{Calabrese2007, Rigol2008}. That hypothesis assumes that the expectation value of a one-particle observable $\mathcal{A}$ in an energy eigenstate of the Hamiltonian $H\ket{M}=\epsilon_M\ket{M}$ equals the thermal average of the corresponding statistical quantity at the mean energy $\mu = \epsilon_M$: $\langle M | \mathcal{A} \ket{M} = \langle A \rangle (\epsilon_M)$.
Note that this is an inherently time independent statement; it assumes that each single eigenvector of the Hamiltonian incorporates statistical signatures and is sufficient to describe thermal behavior. A possible initial nonstatistical behavior appears as the result of a coherent superposition of eigenstates which dephases with time.

\subsection{Examination of quenched quantum systems}
In the following we will discuss a translationally invariant closed quantum system from first principle. 
The conservation of static quantities like energy, momentum, particle number constrain its dynamics; for any initial state it is deterministically determined at any time in the past or in the future by the Schr\"odinger equation.  However, at zero time and by external influence, the Hamiltonian is changed abruptly. Additionally to a time independent (noninteracting) part $H_0$ a two-particle interaction term  $H_{\text{int}}$ is switched on instantly such that 
\begin{equation}
\label{QQS_def}
H(t) = H_0 + \Theta(t) H_{\text{int}}
\end{equation}
$\Theta(t)$ is the Heaviside step function.
To avoid trivialities, we assume that  $[H_0,H_{\text{int}}] \neq 0$. This implies that the eigenbasis of the Hamiltonian changes at zero time: From the eigenvectors of $H_0$, $\lbrace \ket{m} \rbrace_m$, to the eigenbasis $\lbrace \ket{M} \rbrace_M$ of the  Hamiltonian $H$. For convenience, we will assume that nondegenerate perturbation theory can be applied; then a weak interaction does not change the noninteracting eigenenergies dramatically such that a relation can be established between the corresponding energies $\epsilon_m \approx \epsilon_M$ for $m=M$.

In general, a quantum quench implies that the system is open for the intake of energy at the quenching time $t=0$ and for a redefinition of its Hamiltonian ground state.   Yet it never means a sudden reset of its quantum state. Therefore, the quench initializes the interacting system described by the Hamiltonian $H(t>0)$ in the ground state of the noninteracting system $\ket{\Omega_0} = \ket{m=0}$ which typically represents an excited state of the interacting system. Its properties can be studied by following its nontrivial successive dynamics which is, again, generated by a time independent Hamiltonian. Hence, effectively, we study an initial value problem for a time independent Hamiltonian $H=H_0 + H_{\rm int}$. This observation determines our technical approach. 

Secondly, we choose a suitable observable to study the resulting dynamics. 
For translationally invariant systems the momentum mode number operator $\mathcal{N}_k$ is a convenient choice. It commutes with the noninteracting Hamiltonian $[\mathcal{N}, H_0]=0$ such that a common eigenbasis of both operators exists. This justifies the application of perturbation theory to study the dynamics of this observable\footnote{For this special situation the evaluation of the matrix element in the right hand side of (\ref{Intro_ETH}) is simplified; it could be approached, essentially, with a resummation of perturbation theory. Since for a fermionic system the eigenvalues of the number operator are bounded between zero and one, the result will depend on the filling.}.
As the momentum mode number operator is a one-particle observable, it only exposes limited information on the interacting quantum system. This makes thermalization of its expectation value a plausible scenario.

\subsection{Solution of the Heisenberg equations of motion for an observable}
As a final part of the introduction, we construct an appropriate method 
to study analytically the effects of a quantum quench on observables. 

Working in a Heisenberg picture, states remain time independent while the operators carry the total time dependence. The dynamics of observables $\O$ which do not explicitly depend on time is described by the Heisenberg equation of motion \cite{Baym_QM}
\be
\label{QHO_HEM}
i \hbar \frac{d\O(t)}{dt} = [\O(t),H] .
\ee
This equation holds as an operator identity independent of any particular choice of a  basis representation. Its solution, however, is best performed in an exact eigenbasis of the Hamiltonian where the dynamics of different energy scales is decoupled and can be treated separately. There time evolution leads to a pure dephasing of the Hamiltonian eigenmodes. 

In the following, we make this observation useful for approximating the time evolution of arbitrary observables in a more sophisticated application of perturbation theory. Applying time dependent perturbation theory directly to an observable typically results in the generation of \emph{secular terms} which are both proportional to time and the perturbative expansion parameter. They eventually spoil the validity of time dependent perturbation theory even for small expansion parameters. Hence the aim is to separate a perturbative treatment of interaction effects from time evolution. This can be achieved by mapping the observable into an eigenbasis of the Hamiltonian where an exact evaluation of time evolution can be achieved. The corresponding transformation is the same which diagonalizes the Hamiltonian. Yet for many problems it is not known exactly. Implementing it by means of a perturbative expansion generates an approximate representation of the observable in an approximate Hamiltonian eigenbasis. Then secular terms arise again but, fortunately, in higher orders of the expansion only, such that the validity of time dependent perturbation theory has been improved. 

On the other hand, we have understood the quench scenario as an initial value problem and a competing requirement follows from the observance of nonequilibrium initial conditions. Note that the later are defined in the eigenbasis of the noninteracting Hamiltonian $H_0$. Particularly in the context of a many-body initial state, their mapping into a different basis representation may be complicated such that their evaluation is only convenient in their initial one.

\begin{figure}
 \begin{center}
  \includegraphics[height=25mm]{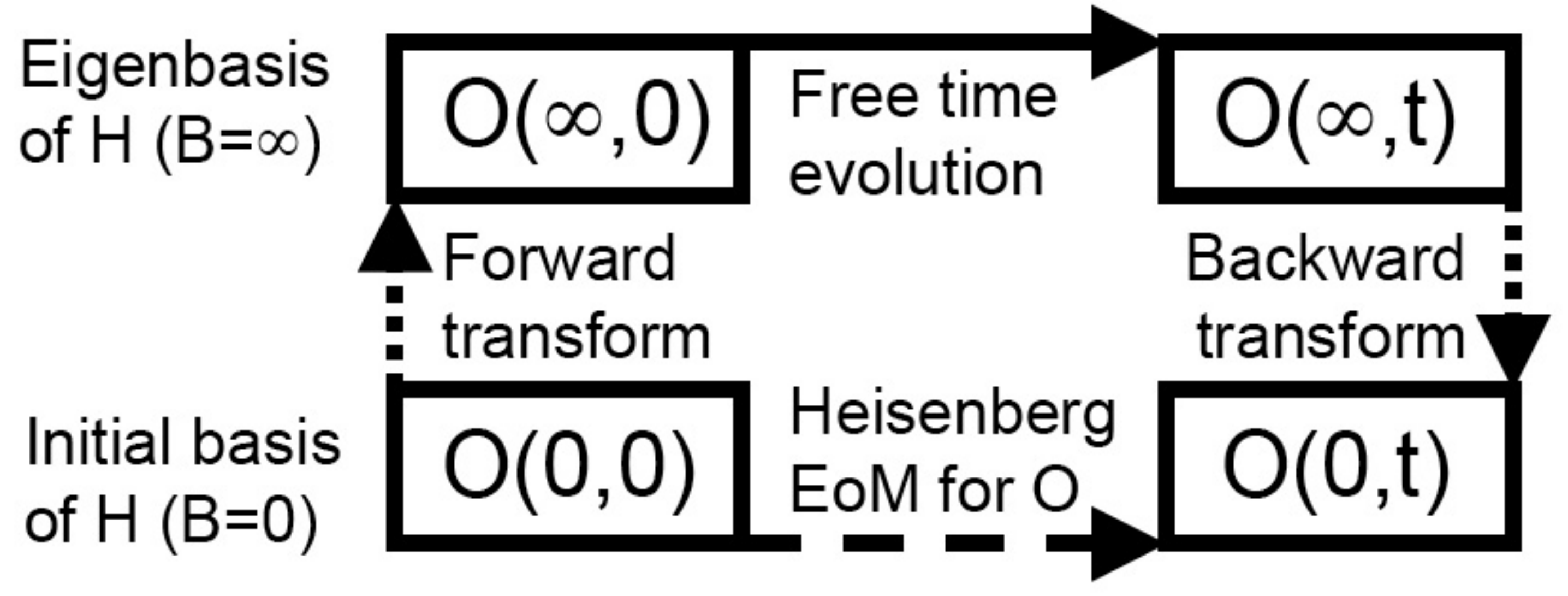} 
    \caption{The Heisenberg equation of motion for an observable $O$ is solved by transforming to 
    the $B=\infty$ eigenbasis of the interacting 
    Hamiltonian $H$ (forward transformation), where the time evolution can be computed easily.
   Time evolution introduces phase shifts, and therefore the form of the observable in the initial
   basis $B=0 $ (after a backward transformation) changes as a function of time.}
  \label{Fig_Rechenschema}
  \end{center}
 \end{figure}

This motivates an approach which combines three unitary transformations: one which diagonalizes the interacting Hamiltonian approximately, called the \emph{forward} transformation, a second one which represents a solution of (\ref{QHO_HEM}) in the interacting eigenbasis and a final \emph{backward} transformation (which reverts the diagonalization) into the noninteracting eigenbasis. All transformations are applied sequentially to the observable and implemented in perturbation theory. Note that the inverse transformation does not reproduce the original structure of the observable since time evolution has dephased the different contributions present in its forward-transformed representation.  This scheme has been described in  \cite{Hackl2007, Hackl2008} and is illustrated by fig. \ref{Fig_Rechenschema}. It is motivated by canonical perturbation theory in classical mechanics.  There the time reliability of a perturbative expansion can be greatly improved if it is performed after a suitable canonical transformation has been applied \cite{Goldstein_CM}.    

\subsection{Outline}
In the following part of this paper we will apply the above approach first to systems with a discrete energy spectrum. 
As an introductory example we discuss 
the exactly solvable model of a suddenly squeezed harmonic oscillator to illustrate technical aspects and the role of perturbative arguments. 
Then we will formulate a more general statement on the relation between equilibrium and nonequilibrium expectation values for certain observables in discrete systems. We give two proofs which highlight different aspects of the characteristic nonequilibrium physics: The first proof stresses the role of perturbative arguments and restrictions imposed on the class of discussed observables by focussing on the overlap of eigenstates of a noninteracting Hamiltonian and its weakly perturbed counterpart. The second proof mirrors the operator approach depicted in fig. \ref{Fig_Rechenschema}. It makes the drop-out of of transient and oscillatory behavior under time averaging more explicit. 

In a second part, we give details on the quench of a Fermi liquid as it has been described in  \cite{Moeckel2008}. This constitutes a many-particle problem with a continuous spectrum of eigenenergies and requires more elaborate diagonalization techniques. Therefore we introduce the flow equation method following  Wegner  \cite{Wegner1994} which is an established approach towards the approximate diagonalization of many-body Hamiltonians and include it into the forward-backward transformation scheme. Although it is a nonperturbative technique we will only use it in an approximate form to set up perturbation theory. This is sufficient to observe the first phase of the dynamics of the quenched Fermi liquid. Studying the momentum distribution around the Fermi surface  mirrors one-particle nonequilibrium physics. In particular, we observe a characteristic nonequilibrium value for the discontinuity of the momentum distribution at the Fermi surface which indicates the size of the quasiparticle residue. In equilibrium, correlation effects lead to a reduction of its value which is one only in the case of the noninteracting Fermi gas. In nonequilibrium, this reduction is twice as large. This factor of two corresponds to the analogous perturbative results for the one-particle squeezed oscillator and in discrete systems. In a Fermi liquid, however, its appearance only indicates a transient nonequilibrium state. This transient behavior exhibits prethermalization  \cite{Berges2005} since contrary to the nonequilibrium momentum distribution the kinetic and the interaction energy have already relaxed to their final long time values.  
The relaxation of the momentum distribution gives rise to a second phase of the dynamics of a quenched Fermi liquid. It will be obtained from a quantum Boltzmann equation which describes the effective evolution of the momentum distribution from the nonequilibrium transient state onwards and leads to the prediction of its thermalization.

\section{Squeezing a one-particle harmonic oscillator}
\label{SecSOPHO}

The squeezed oscillator is a well-studied one-particle model system which found appreciation in many branches of physics. For two decades researchers have discussed squeezed states of the electromagnetic field which are interesting because of their characteristic reduced fluctuations in one field quadrature as compared to coherent states  \cite{Walls1983, Scully_QO}. 
This suppression of quantum fluctuations in one variable out of a set of non-commuting variables below the threshold obtained for a state of symmetrically distributed minimal uncertainty, i.e. a coherent state, has motivated the naming: In this parameter the phase space portrait of the squeezed state shows sharp details and appears 'squeezed' when compared to that one of a coherent state while fluctuations are inevitably increased in the others. 
The physical relevance of squeezed states in optics is grounded on the fact that some interesting phenomena like gravitational waves generated in astronomical events are characterized by oscillations with amplitudes close to or below the width of the ground state wave function of an optical light mode as it is required by Heisenberg's uncertainty principle. Squeezed states, however, may provide improved signal-to-noise ratios beyond this quantum limit of coherent light and simplify the interferometric detection of the weak signatures of gravitational waves  \cite{Goda2008}.

It has been shown \cite{Lo1990} that squeezed states cannot be generated adiabatically from the ground state of a quantum mechanical oscillator but that sudden changes have to be applied to its parameters, e.g. its frequency or spring constant. Therefore, squeezed state represent an early example of what is now, in the context of a many-body system, called a quench of a quantum system.

In many-body theory, the squeezing operation comes under the name of a Bogoliubov transformation. Recently, it was applied to study the behavior of a quenched Luttinger liquid in terms of bosonic degrees of freedom  \cite{Cazalilla2006}. Here we use it to illustrate characteristic nonequilibrium behavior of one-particle models. Since the bosonic representation of a Luttinger liquid is momentum diagonal it equally serves as an example of effective one-particle behavior.

\subsection{Hamiltonian}
On the level of the Hamiltonian, squeezing is inferred by an instantly applied change of the prefactor of the quadratic potential, namely the spring constant. 
We neglect a linear shift of the potential minimum and reduce squeezing to a sudden switch in the coupling constant $g(t)=g \Theta(t) $ of the quadratic particle non-conserving operators. With $\hbar =\omega_0=m=1$
\begin{equation}
\label{QHO_Ham}
H = H_0 + H_{\text{int}}, \  H_0 = a^{\dagger} a + \frac 12,  \ H_{\text{int}} = g(t) \left( (a^{\dagger})^2 + a^2 \right)
\end{equation}
Representing the Hamiltonian in terms of space $x = (a^{\dagger} +a)/\sqrt{2}$ and momentum $p=(a-a^{\dagger})/\sqrt{2}$ operators 
\be
H = \left( \frac 12 +g(t) \right) x^2 + \left( \frac 12 -g(t) \right) p^2 
\ee
shows that it is strictly positive for $\abs{g}\leq1/2$ and, thus, it is bounded from below. We only discuss the Hamiltonian for $0\leq g \leq 1/2$.  

In the following we compare a perturbative analysis of this quench with an exact solution based on the exact diagonalization of $H$.  In both cases we calculate the occupation, i.e. the expectation value of the number operator $\hat{N}=c^{\dagger}c$ both in the equilibrium ground state of the interacting Hamiltonian $H$ and as a long-time limit of the dynamics of an initial state. We will show a remarkable relation between the equilibrium result and the nonequilibrium result which are described by the same functional dependence but variant prefactors. These nontrivial prefactors will later play a key role and can already be appreciated in this simple system. 

\subsection{ Perturbative study of squeezing}
\label{QHO_PD}
Let us first assume that the coupling $g$ is a small parameter and that nondegenerate perturbation theory can be applied to the interacting Hamiltonian. We formulate the perturbative approach in an operator language which corresponds to the formalism of the flow equation method. 

\subsubsection{Definition of the diagonalizing transformation}
\label{QHO_SEC_Trafo}
The first step is to implement a \emph{discrete} unitary transformation which diagonalizes the Hamiltonian to leading order in $g$. We represent the unitary transformation $U(\varphi)=e^{-\eta \varphi}$ by its generator $\eta$, which is an anti-Hermitian operator, and by a scalar angle variable $\varphi$. Then the action of the transformation onto the Hamiltonian can be expanded according to the Baker-Hausdorff-Cambell relation as
\begin{equation*}
\tilde{H} \equiv U^{\dagger}H U= e^{\eta \varphi} H e^{-\eta \varphi}  \approx 
H_0 + H_{\text{int}} + \left( [\eta, H_0] + [\eta, H_{\text{int}}] \right)\varphi  + \frac 12 [\eta,[\eta,H_0]] \varphi^2 + \ldots
\end{equation*}

Demanding that to leading order the interaction term should vanish leads to an implicit definition of the generator 
\be
\label{SHO_eta_impldef}
\varphi [\eta, H_0] = -H_{\text{int}}.
\ee
Note that it implies $\eta \sim \Or(g)$.
It can be easily checked that the \emph{canonical generator} defined as the commutator of the noninteracting and the interacting part of the Hamiltonian fulfills this implicit definition (\ref{SHO_eta_impldef}) if an angle $\varphi = 1/4$ is chosen: 
\be
\label{SHO_eta_can_gen_def}
\eta = \commutator{H_0}{H_{\text{int}}} = 2g  \left( (a^{\dagger})^2 - a^2 \right)
\ee

\subsubsection{Transformed Hamiltonian}
In a second step we consider the corrections beyond leading order in the (approximately) diagonalized Hamiltonian which are, in general, second order in $g$. 
\begin{eqnarray*}
\tilde{H} &=& H_0+ [\eta, H_{\text{int}}] \varphi + \frac 12  [\eta,[\eta,H_0]] \varphi^2
= H_0 - 16 \varphi g^2 H_0 + 32 \varphi^2 g^2 H_0  + \Or(g^3) 
\\ &\stackrel{\varphi = 1/4}{=}& (1-2g^2)H_0 +  \Or(g^3)
\end{eqnarray*}
Thus the transformed Hamiltonian is described by a renormalized frequency \mbox{$ \omega = (1-2g^2)$}\footnote{We point out that this equation constitutes the discrete analogue of a flow equation for the Hamiltonian (cf. \ref{FlowEquO}).}. Due to the particular simplicity of squeezing a harmonic oscillator the second order correction can be fully absorbed in a renormalization of parameters.

\subsubsection{Transformation of quantum mechanical observables}
\label{QHO_SEC_T_obs}
Similarly to its action onto the Hamiltonian the unitary transformation implies a transformation of all quantum mechanical observables which constitutes the third step of a unitary diagonalization approach. 
\be
\label{QHO_BT_TGL}
\tilde{\O} =  \O + \commutator{\eta}{\O}\varphi + \frac 12 \commutator{\eta}{\commutator{\eta}{\O}} \varphi^2 + \ldots
\ee
We make the transformation of creation and annihilation operators explicit up to second order in $g$ and write 
\be
\label{QHO_ForwTrafo_Res}
\left( \begin{array}{c} \tilde{a}^{\dagger} \\ \tilde{a} \end{array} \right) =
\left( \begin{array}{cc} 1+g^2/2 & -g \\ -g & 1+g^2/2 \end{array} \right) 
\left( \begin{array}{c} {a}^{\dagger} \\ {a} \end{array} \right) 
=:T(g)
\left( \begin{array}{c} {a}^{\dagger} \\ {a} \end{array} \right) 
\ee

The three steps (\ref{QHO_SEC_Trafo} - \ref{QHO_SEC_T_obs}) establish a diagonal representation and are, altogether, referred to as the forward transformation. 
We note that the transformation of the observables can be easily inverted. Up to second order in $g$ the inverse of $T$, called the \emph{backward transformation}, is given as $T^{-1}(g) = T(-g)$.

\subsubsection{Spin-off: The equilibrium occupation}
We interrupt our calculation of the nonequilibrium occupation for a short detour in order to evaluate the equilibrium one. To be more specific, our interest is in the occupation of the interacting ground state with 'physical' particles (i.e. particles which are defined by the eigenmodes of the interaction-free Hamiltonian $H_0$). Denoting the interacting ground state by $\ket{\Omega}$ the equilibrium occupation reads $N^{\rm EQU} = \bra{\Omega} a^{\dagger}a\ket{\Omega}$. This is unitarily equivalent to the evaluation of a transformed number operator with respect to the noninteracting ground state $\ket{\Omega_0}$ since $N^{\rm EQU} =\bra{\Omega} \mathcal{U}^{\dagger}\mathcal{U}a^{\dagger}a\mathcal{U}^{\dagger}\mathcal{U}\ket{\Omega}=\bra{\Omega_0} \mathcal{U}a^{\dagger}a\mathcal{U}^{\dagger}\ket{\Omega}$. Fortunately, the transformation which links both representations is the inverse of the forward transformation. Hence we evaluate with $n_a \equiv \bra{\Omega_0} a^{\dagger} a \ket{\Omega_0}$ up to second order in $g$
\begin{eqnarray}
\nonumber
N^{\rm EQU} &=& \bra{\Omega_0} \tilde{a}^{\dagger} \tilde{a} \ket{\Omega_0} 
\approx \nonumber
\left(1+ \frac {g^2}{2}\right) ^2 \bra{\Omega_0} a^{\dagger} a \ket{\Omega_0}  + g^2  \bra{\Omega_0}  a a^{\dagger}  \ket{\Omega_0}  
 \\&\approx& 
(1 +  2 g^2) n_a + g^2 
\label{QHO_BT_N_EQU}
\end{eqnarray}
We note that the occupation of the oscillator measured in terms of the original 'particles' is increased. In the following we will compare this result with the nonequilibrium occupation obtained after sudden squeezing.

\subsubsection{Time evolution of transformed observables}
We resume the calculation for the nonequilibrium case. The forward transformation has already been completed in (\ref{QHO_SEC_Trafo} - \ref{QHO_SEC_T_obs}). In a fourth step the transformed observables are time evolved for all positive times with respect to the transformed Hamiltonian. This, effectively, accounts for the insertion of time dependent phase factors. 
\begin{multline*}
\nonumber
\left( \begin{array}{c} \tilde{a}^{\dagger}(t) \\ \tilde{a}(t) \end{array} \right) =
\left( \begin{array}{c}  e^{i H t} \tilde{a}^{\dagger} e^{-iHt}\\ e^{i H t} \tilde{a} e^{-iHt} \end{array} \right) =
\left( \begin{array}{c}  e^{i \omega t} \tilde{a}^{\dagger} \\ e^{-i \omega t} \tilde{a} \end{array} \right) \\
\stackrel{(\ref{QHO_ForwTrafo_Res})}{=}
\left( \begin{array}{cc} e^{i\omega t}(1+g^2/2) & -e^{i\omega t} g \\ - e^{-i\omega t} g & e^{-i\omega t}(1+g^2/2) \end{array} \right) 
\left( \begin{array}{c} {a}^{\dagger} \\ {a} \end{array} \right) 
\end{multline*}

\subsubsection{Backward transformation}
Finally, we map back the time-evolved observables to the eigenbasis of the noninteracting Hamiltonian, completing the scheme depicted in fig. \ref{Fig_Rechenschema}. Up to second order in $g$ we obtain
\begin{multline*}
\left( \begin{array}{c} {a}^{\dagger}(t) \\ {a}(t) \end{array} \right) =
T^{-1}(g) \left( \begin{array}{c}  \tilde{a}^{\dagger}(t) \\ \tilde{a}(t) \end{array} \right) = \\
\left( \begin{array}{cc} 
e^{i\omega t} + 2i g^2 \sin(\omega t) & -2ig(1+g^2/2) \sin (\omega t) \\ 
2ig(1+g^2/2) \sin (\omega t) & e^{-i\omega t}-2ig^2 \sin(\omega t) \end{array} \right) 
\left( \begin{array}{c} {a}^{\dagger} \\ {a} \end{array} \right) 
\end{multline*}
This constitutes a consistent perturbative solution of the Heisenberg equations of motion for the operators $a^{\dagger}$ and $a$. 

\subsubsection{Nonequilibrium occupation}
In a final step we compose the time dependent number operator from the time dependent creation and annihilation operator in an obvious way. Since time evolution is unitary, the time evolution of a product of operators is always the product of the time evolved operators which can be easily checked by inserting unity $\unity = U(t,t_0)U^{\dagger}(t,t_0)$. 
Evaluating the expectation value of the number operator for the initial state $\ket{\Omega_0}$ leads to the nonequilibrium occupation
\begin{eqnarray}
N^{\rm NEQ}(t) &=& \bra{\Omega_0} a^{\dagger}(t) a(t) \ket{\Omega_0} 
 = 
n_a + 4 g^2 (2 \sin^2 (\omega t) ) (\bra{\Omega_0} a^{\dagger} a \ket{\Omega_0} + \frac 12) \qquad
\end{eqnarray}
The large time limit is obtained by time averaging which is defined for a time dependent variable $A(t)$ as $\overline{A}:=\frac{1}{T} \int_0^T dt A(t)$.  Then \mbox{$\overline{N}^{\rm NEQ} = n_a + 4 g^2 n_a + 2 g^2$.} Comparing with (\ref{QHO_BT_N_EQU}), we find with $\Delta N(t) := N(t) -n_a$ 
\be
\lim_{t \rightarrow \infty} \ \overline{ \Delta N^{\rm NEQ}(t)} \ = \ \mathbf{2} \  \Delta N^{EQU} + \Or{g^3}
\ee
The factor of two between the equilibrium and the nonequilibrium occupation constitutes the main result of this calculation.  It states that even in a long-time limit the nonequilibrium occupation does not approach the equilibrium one. The numerical value of two can be considered as a consequence of applying two transformations, the forward and the backward one, such that changes to the occupation due to interaction effects double. In the following we will find that, although the numerical value gets corrections in order $g^3$, the mismatch of both occupations is retained for all orders of perturbation theory. 

\subsection{Nonperturbative (Bogoliubov) treatment of squeezing}
In a second approach, we implement a Bogoliubov transformation which exactly diagonalizes the squeezing Hamiltonian. It can be found in many textbooks, e.g.  \cite{Scully_QO, Schleich2001}. Our aim is to illustrate that the chosen perturbative approach exhibits, up to numerical details, the correct nonequilibrium behavior of the system. 

The exact diagonalization of the squeezing Hamiltonian (\ref{QHO_Ham}) can be constructed from the action of the (inverse) unitary squeezing operator 
\begin{equation*}
S(\xi) = e^{1/2 \xi^*a^2 - 1/2 \xi (a^{\dagger})^2} 
\end{equation*}
where $\xi = r e^{i\theta}$ is an arbitrary complex number which will be specified later. Applying the squeezing operator to the ground state generates squeezed states in analogy with the displacement operator which maps the ground state onto coherent states. Here we go the opposite way and apply its inverse to diagonalize the squeezing Hamiltonian.

\subsubsection{Exact transformation of observables} 
We directly start with writing down the action of  $S(\xi)$ onto the creation operator and the annihilation operator. In condensed matter theory it is commonly known as a \mbox{Bogoliubov} transformation used to treat interactions quadratic in creation or annihilation operators. 
\begin{equation}
\label{QHO_Bog_OT}
\left( \begin{array}{c} \tilde{a}^{\dagger} \\ \tilde{a} \end{array} \right) =
S^{\dagger}(\xi) \left( \begin{array}{c} a^{\dagger} \\ a \end{array} \right) S(\xi) =
\left( \begin{array}{cc} \cosh(r) & -e^{-i\theta}\sinh(r) \\ -e^{i\theta} \sinh(r) & \cosh(r) \end{array} \right) 
\left( \begin{array}{c} {a}^{\dagger} \\ {a} \end{array} \right) 
=:T^F(\xi)
\left( \begin{array}{c} {a}^{\dagger} \\ {a} \end{array} \right) 
\end{equation}
Note that $T^F(\xi)$ is not a unitary matrix despite the fact that $\det(T^F(\xi)) = 1$. 

\subsubsection{Exact Hamiltonian diagonalization}
Inserting this transformation into the interacting Hamiltonian (\ref{QHO_Ham}) results in a sum of four terms:
\begin{eqnarray*}
\nonumber
\tilde{H} = 
 & (\tilde{a}^{\dagger})^2 &\times\left[ e^{i\theta} \cosh(r) \sinh(r) + 
g \cosh^2(r) + g e^{2i\theta}\sinh^2(r) \right] + h.c. +
\\ &
\nonumber
\tilde{a}^{\dagger} \tilde{a} & \times\left[ \cosh^2(r) + \sinh^2(r)
+ 4 g \cos(\theta) \cosh(r) \sinh(r) \right] +
\\ &\unity&
\times\left[ 2 \cos(\theta) \cosh(r) \sinh(r) \right]
\end{eqnarray*}
To achieve a diagonal Hamiltonian we demand that the terms quadratic in $\tilde{a}^{\dagger}$ and $\tilde{a}$ should vanish. This fixes the free parameter $\xi := r e^{i \theta} = r$. For small interactions $\abs{g}\leq 1/2$ real solutions with $\theta=0$ can be found. With $\sinh(2r) = 2 \sinh(r)\cosh(r)$, $\cosh(2r) = \sinh^2(r)+\cosh^2(r)$ the real parameter $r$ can be linked to the interaction 
\begin{equation*}
\tanh(2r) = -2g 
\end{equation*}
For small values of $g\ll 1/2$ the expansion $\text{arctanh}(x)\sim x$ for $x\ll 1$ implies that $r \approx -g$.
Then the nonperturbative Bogoliubov transformation coincides with the perturbative approach in (\ref{QHO_PD}), with $S(\xi(g)\approx \hspace{-2pt}-g) = \left.e^{\eta(g) \varphi}\right|_{\varphi=1/4}$. 
The diagonal Hamiltonian shows a renormalized frequency $\omega = \cosh(2r) + 2g \sinh(2r)$ compared to the original frequency $\omega_0=1$ in (\ref{QHO_Ham}). 
For all values of $g<1/2$ the renormalized frequency is positive and the Hamiltonian is bounded from below. Its dependence on $g$ is plotted in fig. \ref{Fig_Faktor2}.  In the limit of small $g$ we find for the renormalized frequency its  perturbative value $\omega = (1 -2g^2)$.

\subsubsection{Exact equilibrium occupation}
Again, we first calculate the expectation value of the equilibrium number operator, using $\cosh^2(r)-\sinh^2(r) =1$
\begin{eqnarray}
\nonumber
N^{\rm EQU} &=& \bra{\Omega} a^{\dagger} a \ket{\Omega}  = 
\bra{\Omega_0} \tilde{a}^{\dagger} \tilde{a} \ket{\Omega_0}  
\\ &=& \nonumber 
 \cosh^2(r)  \bra{\Omega_0} {a}^{\dagger} {a} \ket{\Omega_0}  + \sinh^2(r)  \bra{\Omega_0}  {a} {a}^{\dagger}  \ket{\Omega_0}  
 \\
 &=& n_a + 2 \sinh^2(r)\left( n_a + \frac12 \right)    
 \label{QHO_NPT_N_EQU}
\end{eqnarray}
Again, the perturbative limit for small $g$ agrees with (\ref{QHO_BT_N_EQU}).

\subsubsection{Exact nonequilibrium occupation}
For the nonequilibrium occupation we solve the Heisenberg equations of motions for the creation and annihilation operators in the (now exact) eigenbasis of the Hamiltonian.  
The forward transformation of these operators is given by (\ref{QHO_Bog_OT}). 
Again, we complete the scheme in fig. \ref{Fig_Rechenschema} and compute the time evolution of the transformed operators with respect to the diagonalized Hamiltonian, i.e. with respect to the renormalized frequency $\omega$. The final backward transformation is given by $T^B(r)=T^F(-r)$. These three steps can be easily denoted as subsequent matrix multiplications:
\begin{eqnarray*}
\nonumber
\left( \hspace{-5pt}\begin{array}{c} a^{\dagger}(t) \\ a(t) \end{array}\hspace{-5pt} \right)\hspace{-5pt} &=&\hspace{-5pt}
T^B(r)
\left( \hspace{-5pt}
\begin{array}{cc}  e^{-i \omega t} &0 \\ 0 & e^{i \omega t} \end{array}
\hspace{-5pt} \right) 
T^F(r)
\left(\hspace{-5pt} \begin{array}{c} {a}^{\dagger} \\ {a} \end{array} \hspace{-5pt}\right) 
\\ &=&
\left( \hspace{-5pt}
\begin{array}{cc}  e^{-i \omega t} \cosh^2(r)-e^{i\omega t} \sinh^2(r) &
i\sin(\omega t) e^{-i\theta}\sinh(2r) \\ 
-i\sin(\omega t) e^{i\theta}\sinh(2r) & -\left(e^{i \omega t} \cosh^2(r)-e^{-i\omega t} \sinh^2(r) \right) \end{array}
\hspace{-5pt} \right) 
\left(\hspace{-5pt} \begin{array}{c} {a}^{\dagger} \\ {a} \end{array} \hspace{-5pt}\right) 
\label{QHO_CAO_TR}
\\&=& \left[
\cos(\omega t) 
\left(\begin{array}{cc} 1&0\\0&-1 \end{array} \right) 
-i \sin(\omega t)
\left(
\begin{array}{cc}   \cosh(2r)&
-e^{-i\theta}\sinh(2r) \\ 
 e^{i\theta}\sinh(2r) &  \cosh(2r) \end{array} \right) \right]
\hspace{-1pt} 
\left(\hspace{-5pt} \begin{array}{c} {a}^{\dagger} \\ {a} \end{array} \hspace{-5pt}\right) 
\end{eqnarray*}
Composing the number operator reads 
\begin{eqnarray*}
N^{\rm NEQ}(t)
&=& \bra{\Omega_0} a^{\dagger}(t) a(t) \ket{\Omega_0} \\ &=& 
\bra{\Omega_0} \left[
 \left(e^{-i \omega t} \cosh^2(r)-e^{i\omega t} \sinh^2(r) \right) a^{\dagger}
+i\sin(\omega t) e^{-i\theta}\sinh(2r) a \right] \times 
\\&& \qquad
\left[ -i\sin(\omega t) e^{i\theta}\sinh(2r) a^{\dagger}  -\left(e^{i \omega t} \cosh^2(r)-e^{-i\omega t} \sinh^2(r) \right) a \right] \ket{\Omega_0}
\end{eqnarray*}
Only the particle number conserving terms $\sim a^{\dagger} a, aa^{\dagger}$ contribute and we arrive at the nonequilibrium occupation
\begin{equation*}
N^{\rm NEQ}(t) =
n_a + \left(2\sin^2(\omega t) \right)  \sinh^2(2r) \left(n_a + \frac 12\right) 
\end{equation*}
Again, the long time limit is taken as a time average. This implies that the renormalization of the frequency does not affect the occupation at late times and may be neglected. 
\be
\label{QHO_BogT_N_NEQ_LT}
\lim_{t \rightarrow \infty} \overline{N^{\rm NEQ}(t)} =n_a + \sinh^2(2r) \left(n_a + \frac 12\right) 
\ee

\subsubsection{Nonperturbative relation between the equilibrium occupation  and the nonequilibrium occupation}
Comparing (\ref{QHO_BogT_N_NEQ_LT}) with (\ref{QHO_NPT_N_EQU})
one observes that for the squeezing Hamiltonian the relation between the equilibrium and nonequilibrium occupation is given by
\begin{equation}
\label{Faktor2Factorm}
m(r(g)):=\frac{\Delta \overline{N^{\rm NEQ}}}{\Delta N^{\rm EQU}} = \frac{\sinh^2(2r(g))}{2\sinh^2(r(g))}
\end{equation} 
\begin{figure}
 \begin{center}
  \includegraphics[height=55mm]{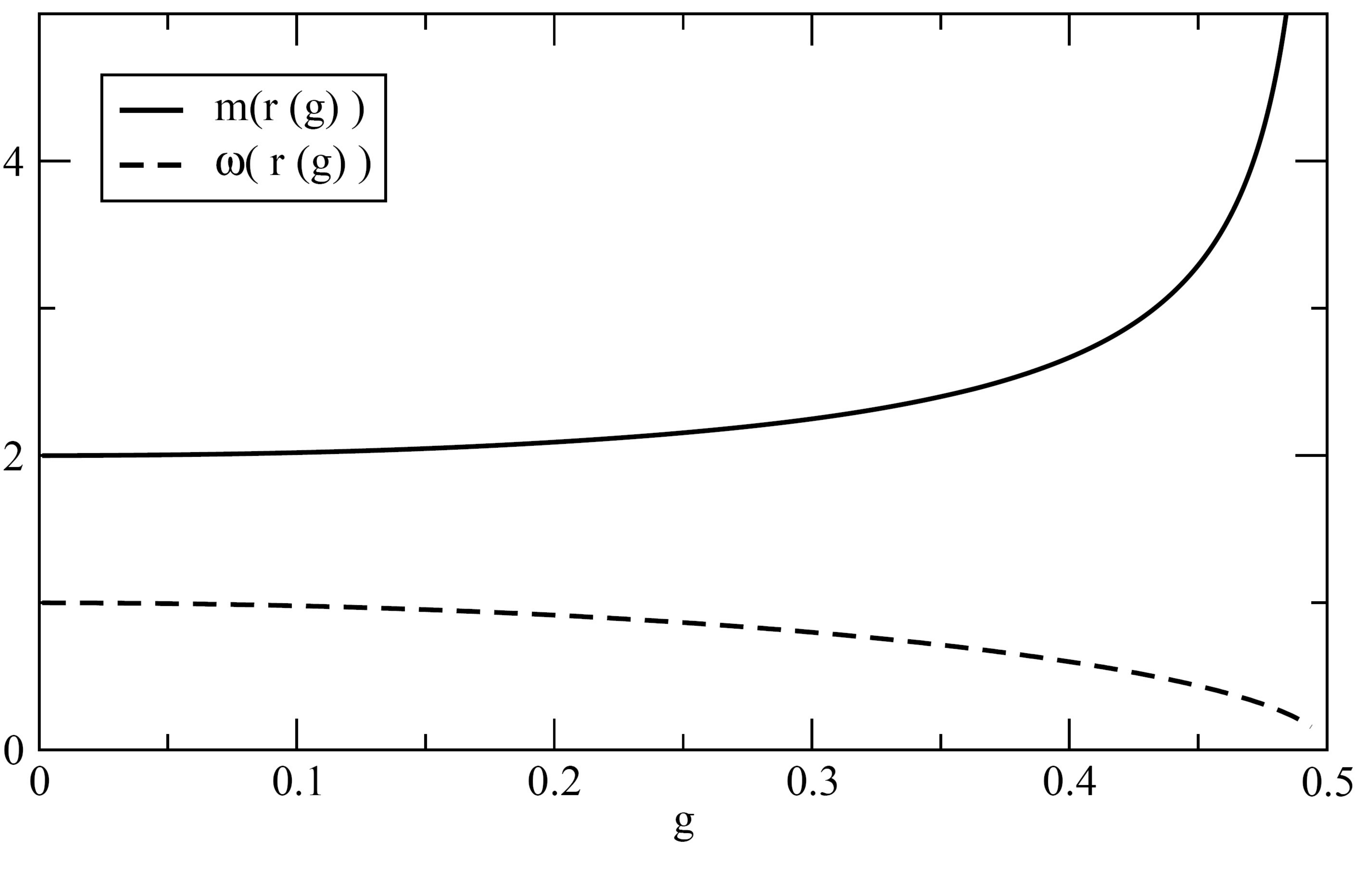} 
    \caption{This plot illustrates the nonperturbative ratio $m(r(g))= \frac{\sinh^2(2r(g))}{2\sinh^2(r(g))}$ in (\ref{Faktor2Factorm}) between the equilibrium and nonequilibrium correction to the noninteracting occupation due to interaction effects. The real solution for the diagonalization transformation $\xi(g)=r(g)$ is valid only for $g < 1/2$. In the limit of small interaction the factor $m=2$ is exact. Additionally the renormalized frequency $\omega(g)$ of the diagonal Hamiltonian is shown. }
  \label{Fig_Faktor2}
  \end{center}
 \end{figure}
 
The precise numeric value of the ratio depends via $r$ on the coupling strength $g$ and is plotted in fig. \ref{Fig_Faktor2}.
Expanding $\sinh(r) \approx r + r^3/3! + \Or(r^5)$ we confirm that in the perturbative limit this relation approaches a factor of two.


\section{More general statements on quenched one-particle systems}
\label{Math_SEC_Proofs}
As the discussion of the squeezed oscillator has shown a perturbative approach to quenched one-particle systems captures important signatures of the behavior of their nonequilibrium occupation. In the following we will show that it can be generalized to a large class of weakly interacting one-particle systems and to more general observables.

\emph{Prerequisites:}
Let us consider the Hamiltonian of a quantum system with a discrete energy spectrum and small interaction $g$ which models a weak quantum quench. 
\be
\label{MG_HAM}
H = H_0 + g \ \Theta(t) H_{\text{int}}
\ee
Its interacting ground state is denoted by $\ket{\Omega}$.  
We assume that nondegenerate perturbation theory with respect to the noninteracting ground state $\ket{\Omega_0}$ is applicable and that $H_0$ and $H_{\text{int}}$ do not commute. 
 
Moreover, we assume a quantum mechanical observable $\O$ which does not depend explicitly on time and obeys the following relations:
\begin{eqnarray}
\O \ket{\Omega_0} &=& 0 \label{Math_Gen_O_Destr} \\
\commutator{\O}{H_0}&=&0 \label{Math_Gen_O_TE}
\end{eqnarray}
For times $t>0$ its time evolution $\O(t)$ is generated by the interacting Hamiltonian $H_0 + g H_{\text{int}}$.
Then the following theorem holds:

\emph {Theorem:}
In second order perturbation theory the long-time limit of the time-averaged expectation value of the time evolved observable in the initial state equals two times the equilibrium expectation value of the observable in the interacting ground state.
\begin{equation}
\label{Theorem}
\overline{\O}:=\lim_{T \rightarrow \infty}\frac{1}{T} \int_0^T dt \ {\bra{\Omega_0} \O(t)\ket{\Omega_0}} 
{=}
2 \bra{\Omega} \O \ket{\Omega} +\Or(g^3)
\end{equation}

\emph{Proof 1:}
We will give two proofs of this theorem. The first one is intended to motivate the physical origins of the prerequisites by relating it to the more conventional picture of overlapping eigenstates. This allows to conclude on its general relevance. 

Firstly, we introduce eigenbasis representations for the noninteracting Hamiltonian $H_0$ $\lbrace \left. \ket{m} \right| m \in N_0\rbrace$\footnote{We label eigenstates of the noninteracting Hamiltonian by \emph{lower} case variables.}, the interacting Hamiltonian $\lbrace \left. \ket{M} \right| M \in N_0\rbrace$\footnote{We label eigenstates of the interacting Hamiltonian by \emph{upper} case variables.} and the observable $\lbrace \left. \ket{j} \right| j \in N_0\rbrace$ with the eigenvalues $\epsilon^0_m=\epsilon_M+\Or(g^2)$ (for $m=M$) and $O_j$, respectively. 
The requirement (\ref{Math_Gen_O_TE}) implies the existence of a common eigenbasis of the observable $\O$ and the noninteracting Hamiltonian $H_0$ such that we can assume pairwise coinciding eigenvectors $\ket{m}=\ket{j}$. For clarity, however, we will keep a separate notation. The equilibrium ground state expectation value is rewritten by inserting unity.
\be
\label{Math_ExV_EQU}
\bra{\Omega} \O \ket{\Omega} = \sum_j O_j \abs{\left\langle j\right.\ket{ \Omega}}^2
\ee
An analogous evaluation of the time dependent expectation value by inserting unities, extracting time dependent phase factors and taking their time average leads to
\begin{multline}
\label{Math_ExV_NEQ}
\overline{\bra{\Omega_0} e^{-iHt} \O e^{iHt} \ket{\Omega_0}}
 =  \\
= \lim_{T\rightarrow \infty} \frac{1}{T} \int_0^T dt 
\sum_{MM'jj'} \bra{\Omega_0} e^{-iHt} \ket{M}\left\langle M\right.\ket{j}\bra{j}\O\ket{j'}
\left\langle j'\right.\ket{M}\bra{M'}e^{iHt} \ket{\Omega_0}
\\  = 
\lim_{T\rightarrow \infty} \frac{1}{T} \int_0^T dt 
\sum_{MM'jj'} O_{j'} e^{i(\epsilon_{M'}-\epsilon_{M})t} \left.\bra{\Omega_0} M\right\rangle\left\langle M\right.\ket{j}\bracket{j}{j'}
\left\langle j'\right.\ket{M}\bracket{M'}{\Omega_0}
\\  = 
\sum_{Mj} O_j   \abs{\left\langle M\right.\ket{ \Omega_0}}^2 \abs{\left\langle j\right.\ket{ M}}^2
\end{multline}
Up to a relative phase, the interacting eigenstates $\ket{M}$ are invariant under time evolution. Therefore, overlap matrix elements are discussed with respect to these states. 

The set of matrix elements $\lbrace\left\langle M\right.\ket{ \Omega_0}\rbrace_M$ describe a decomposition of the initial state in terms of Hamiltonian eigenstates. This is a statement about the particular initial conditions of the quench problem. Since we discuss a quench from the noninteracting Hamiltonian this is a decomposition of the noninteracting ground state in terms of interacting eigenstates.

The second set of matrix elements $\lbrace\left\langle j\right.\ket{M}\rbrace_{M,j}$ encapsulates the overlap between the eigenbasis of the observable and the eigenbasis of the Hamiltonian. 
Since (\ref{Math_Gen_O_TE}) holds we can work in the common eigenbasis of the observable and the noninteracting Hamiltonian. Then the overlap between the eigenbasis of the interacting and of the noninteracting Hamiltonian is discussed. In both cases the matrix elements can be evaluated by applying perturbation theory to the Hamiltonian $H=H_0+H_{\rm int}$, treating $H_{\rm int}$ as a small perturbation.  
We make this explicit to leading order: 
\be
\nonumber
\abs{\left\langle m\right.\ket{M}}^2 \stackrel{\rm PT}{=} \left \lbrace
\begin{array}{cc} 1& \text{for } M=m \\ \left|\frac{\bra{m} H_{\rm int} \ket{M}}{(\epsilon^0_{M} - \epsilon_{m}^0)}\right|^2 & \text{for }  M \neq m
\end{array} \right.
\ee
As (\ref{Math_Gen_O_Destr}) implies $O_0=0$ the direct overlap between the interacting and the noninteracting ground state does not contribute to the sums in both (\ref{Math_ExV_EQU}) and (\ref{Math_ExV_NEQ}); hence they are at least second order in $g$. We compare the right hand side of both equations for any fixed value of $j$.
In the nonequilibrium case, second order contributions require a resonance condition for the involved quantum numbers, $M=j$ or $M=0$. 
\be
\nonumber
\abs{\left\langle M\right.\ket{ \Omega_0}}^2 \times \abs{\left\langle j\right.\ket{ M}}^2 \stackrel{\rm PT}{=} \left \lbrace
\begin{array}{cc} 
\abs{\left\langle J\right.\ket{ \Omega_0}}^2 \times 1& \text{for } M=j=:J \\ 
1 \times \abs{\left\langle j\right.\ket{ \Omega}}^2  & \text{for }  M = 0
\end{array} \right.
\ee
Because of the symmetry $\abs{\left\langle J\right.\ket{ \Omega_0}}^2 = \abs{\left\langle j\right.\ket{ \Omega}}^2$ in leading order perturbation theory, both contribute equally $\abs{\left\langle j\right.\ket{ \Omega}}^2$ to the sum over $M$. Then in second order perturbation theory holds 
\be
\nonumber
\overline{\bra{\Omega_0} e^{-iHt} \O e^{iHt} \ket{\Omega_0}}=2 \sum_j \O_j \abs{\left\langle j\right.\ket{ \Omega}}^2 = 2 \bra{\Omega} \O \ket{\Omega} 
\ee
and the theorem is proven.

\emph{Proof 2:}
The second proof of the theorem is constructed in analogy to the scheme presented in fig. \ref{Fig_Rechenschema} and aims at a clearer understanding of the applied method and its particular merits. 

\emph{ (1) Definition of a unitary transformation.} We define a \emph{single} unitary transformation $\U_s^{\dagger}= e^{\eta_s}$ by its anti-hermitian generator $\eta_s=-\eta_s^{\dagger}$, demanding that its application to the Hamiltonian disposes the interaction part of the Hamiltonian to first order of $g$. Expanding its unitary action onto the Hamiltonian  
\begin{equation}
\tilde{H} = e^{\eta_s} H e^{-\eta_s} = 
H_0 + \underset{\Or(g)}{\underbrace{H_{\text{int}}+\commutator{\eta_s}{H_0}}}
+\commutator{\eta_s}{H_{\text{int}}}
+\label{MG_PR_H_Expanded}
\frac 12 \commutator{\eta_s}{\commutator{\eta_s}{H_0+H_{\text{int}}}} + \Or(g^3)
\end{equation}
hence allows to read off an implicit definition of $\eta_s$ by 
$$ \commutator{\eta_s}{H_0}{=}-H_{\text{int}} $$ 
which justifies the assumption $\eta_s \sim \Or(g)$ in (\ref{MG_PR_H_Expanded}). Then the transformed Hamiltonian equals the free Hamiltonian up to second order corrections.

\emph{(2) Computation of the interacting ground state expectation value.}
In the following we exploit a formal coincidence which holds for all systems with a nondegenerate single ground state: For any such Hamiltonian, the diagonal representation of the interacting ground state in terms of the diagonal degrees of freedom can be formally identified with the ground state of the noninteracting Hamiltonian; thus  we can relate them by $\U_s^{\dagger}\ket{\Omega}=\ket{\Omega_0}$ or, to leading order, by
$\ket{\Omega} = 
(1-\eta_s ) \ket{\Omega_0}+\O(g^2)$.
As every Hamiltonian can be diagonalized, this does not pose any further restrictions. Hence with (\ref{Math_Gen_O_Destr})
\begin{equation}
\bra{\Omega} \O \ket{\Omega} 
= \bra{\Omega_0} \U_s^{\dagger}\O \U_s \ket{\Omega_0} 
{=} -\bra{\Omega_0} \eta_s\O\eta_s \ket{\Omega_0} + \Or(g^3)  
\label{Math_Gen_IGS_EV}
\end{equation}
The simple diagonal representation of the interacting ground state motivates the application of operator-based transformation schemes like the flow equation method in equilibrium since correlation effects are, formally, fully transferred from the description of an interacting ground state to the the particular form of transformed observables. Thus one can avoid to discuss the full complexity of the interacting ground state and restrict to those correlation effects which become actually relevant for a particular observable. The transformation $\U_s^{\dagger} \O \U_s$ can be performed in the most convenient way. 

\emph{(3) Real-time dynamics of the observable after the quench.}
For the evaluation of the nonequilibrium expectation value $\overline{\O}$ we start with the sequential application of three unitary transformations. Firstly, at time $t=0$ the observable is represented approximately in the energy-diagonal eigenbasis of the Hamiltonian. 
\be
\label{Math_OT_FT}
\tilde{\O}(0) = \O(0) + \commutator{\eta_s}{O(0)} + \frac 12 \commutator{\eta_s}{\commutator{\eta_s}{\O(0)}} + \Or(g^3)
\ee
Now we apply unitary time evolution to the transformed observable with respect to $\tilde{H}= H_0+\Or(g^2)$. This is time-dependent perturbation theory to first order. 
\be
\label{Math_OT_TE}
\tilde{\O}(t)= e^{-i{H}_0t} \tilde{\O}(0) e^{i{H}_0t}
\ee
We insert (\ref{Math_OT_FT}) into (\ref{Math_OT_TE}) and attribute the time dependence to the generator $\eta_s \stackrel{}{\rightarrow} \eta_s(t)=e^{-iH_0t}\eta_s(0) e^{iH_0t}$. This is possible because of (\ref{Math_Gen_O_TE}) and ensures that (\ref{Math_Gen_O_Destr}) holds for all times.
Finally, the backward transformation $\eta_B = - \eta_s(0)$ is applied.
\begin{multline*}
\O(t) = \tilde{\O}(t) - \commutator{\eta_s(0)}{\tilde{\O}(t)} + \frac 12
\commutator{ \eta_s(0)}{\commutator{\eta_s(0)}{\tilde{\O}(t)}} + \Or(g^3)\\
=
{\O(0)} + \commutator{\eta_s(t)}{{\O}} + \frac 12
\commutator{ \eta_s(t)}{\commutator{\eta_s(t)}{\O}}
 - \commutator{\eta_s(0)}{\O} \\  - 
\commutator{ \eta_s(0)}{\commutator{\eta_s(t)}{\O}} +
\frac 12 \commutator{ \eta_s(0)}{\commutator{\eta_s(t)}{\O}} + \Or(g^3)
\end{multline*}
We evaluate the expectation value of $\O(t)$ in the initial state $\ket{\Omega_0}$. Due to (\ref{Math_Gen_O_Destr}) many contributions vanish.
\begin{multline}
\bra{\Omega_0} \O(t)\ket{\Omega_0} = 
\left \langle 2 \left(-\frac 12 \right)  \eta_s(t) \O \eta_s(t) +
\eta_s(0) \O \eta_s(t) \right. 
\\ 
\left.
+ \eta_s(t) \O \eta_s(0)- 2 \left(\frac 12 \right) \eta_s(0) \O \eta_s(0) \right\rangle_{\Omega_0}
\label{Math_Gen_RTE_O}
\end{multline}
Inserting unity $\unity = \sum_m \ket{m}\bra{m}$ in terms of eigenstates of the noninteracting Hamiltonian $H_0$ shows that the second and the third term in (\ref{Math_Gen_RTE_O}) dephase and do not contribute to the long time average:
\begin{eqnarray}
\bra{\Omega_0}\eta_s(0) \O \eta_s(t) \ket{\Omega_0} 
&=&
\sum_m \bra{\Omega_0}\eta_s(0) \O \ket{m}\bra{m}e^{-iH_0t}\eta_s(0) e^{iH_0t} \ket{\Omega_0}  
\\ &=&
\sum_m O_m e^{i(\epsilon_m-\epsilon_0)t}\bra{\Omega_0}\eta_s(0) \ket{m}\bra{m}\eta_s(0) \ket{\Omega_0} 
\end{eqnarray}
For $m=0$ equation (\ref{Math_Gen_O_Destr}) implies $\O_0=0$. As we have assumed a nondegenerate Hamiltonian $H_0$ we obtain 
\be
\nonumber
\overline{\bra{\Omega_0}\eta_s(0) \O \eta_s(t) \ket{\Omega_0}}=\overline{\bra{\Omega_0}\eta_s(t) \O \eta_s(0) \ket{\Omega_0}}=0
\ee
On the other hand, making use of  (\ref{Math_Gen_O_TE}) 
\begin{equation*}
\bra{\Omega_0} e^{-iH_0t} \eta_s e^{iH_0t} \O e^{-iH_0t} \eta_s e^{iH_0t} \ket{\Omega_0} = \bra{\Omega_0} \eta_s(0) \O  \eta_s(0) \ket{\Omega_0}
\end{equation*}
Consequently, we arrive at
$
\overline{\O} =- 2 \bra{\Omega_0} \eta_s(0) \O \eta_s(0) \ket{\Omega_0}
$.
With (\ref{Math_Gen_IGS_EV}) the theorem is proven.

The second proof explains the factor of two as the accumulation of equal second order corrections both from the forward and from the backward transformation. The drop-out of transient or oscillatory behavior in (\ref{Math_Gen_RTE_O}) due to time averaging is more explicit. This depicts the major merit of the transformation scheme: Fundamental correlation-induced effects -- as it is, for example, the difference between the interacting and the noninteracting ground state-- enter a perturbative study of time evolution performed in an energy-diagonal representation already as time-independent offsets.  That their influence is stronger in nonequilibrium than in equilibrium can be seen directly. 

\emph{Corollary:} In many systems the noninteracting part $H_0$ of an interacting Hamiltonian $H$ represents the kinetic energy for which the following relation holds:\footnote{The increase in the kinetic energy beyond its value for the interacting ground state, however, does not necessarily indicate its thermal distribution (i.e. heating effects).}
$\overline{E^{\text{NEQ, KIN}}}=  2 E^{\rm EQU, KIN}$.

\emph{Proof:} Define $E^{\text{NEQ, KIN}}(t):=\bra{\Omega_0} H_0(t) \ket{\Omega_0}$ and apply the theorem for $\O=H_0$.

\subsubsection*{Discussion:}
The above theorem explains the factor 2 in the ratio between nonequilibrium and equilibrium expectation values as a rather general observation in systems with discrete energy spectra. In the following we will see how this factor 2 appears for an interaction quench in a Fermi liquid with continuous spectrum and what role it plays for the nonequilibrium dynamics. Notice that recently similar observations have been made for a quenched Kondo impurity in the ferromagnetic regime\cite{Hackl2009}. 
There the interaction of the impurity spin with a band of metal electrons is switched on suddenly in time and the subsequent spin dynamics is calculated. The nonequilibrium dynamics of the magnetization has been studied by both analytical (flow equations) and exact numerical (time dependent NRG) methods\cite{Hackl2009}. The respective results agree very well on all time scales and again show the above factor 2 when comparing to the equilibrium value. 


\section{Real-time evolution of a quenched Fermi liquid}


In the last section we have compared the equilibrium and nonequilibrium behavior of a single-particle anharmonic oscillator. In the following we present an analogous analysis for a many-body noninteracting Fermi gas for which we compare a sudden switching-on of a two-particle interaction with an adiabatic evolution into an interacting ground state. In more than one dimension the later corresponds to Landau's theory of a Fermi liquid, which, since its introduction in the late 1950s  \cite{Landau1957a}, became a benchmarking effective description for the study of many (normal) interacting Fermi systems  \cite{Baym_FLT, Pines1966}. 
Its main prerequisite (at least from the point of view of this work), is the \emph{adiabatic connection} between the noninteracting Fermi gas and the interacting Fermi liquid. It means that there is a continuous evolution from the low energy states of the gas to those of the liquid as the interaction is increased. This continuous link allows to formulate the intuitively appealing concept of \emph{Landau quasiparticles} since particle properties carry over from the noninteracting physical fermions to the interacting degrees of freedom: Around the Fermi surface, the interacting degrees of freedom, then called quasiparticles, differ from the noninteracting ones only by modified parameters, e.g.  an effective mass. In terms of physical fermions Landau quasiparticles are composite many-particle objects although they are not true eigenstates of the system; this implies both a residual interaction among them and their finite lifetime which roughly measures the departure from interacting eigenstates. As the lifetime diverges right at the Fermi energy Landau's approach becomes exact there. Away from the Fermi energy two incompatible time scales compete with each other, limiting the applicability of Landau's approach to low energy excitations:  The adiabatic requirement of long ramp-up times and the limited lifetime of the achieved quasiparticle picture. 
In this work we will discuss the opposite limit of the adiabatic increase of the interaction strength by applying a sudden quench. We will find that for weak quenches within the Fermi liquid regime a quasiparticle picture can be retained. 

\subsection{Hubbard model}

We discuss a generic Fermi liquid by referring to the Fermi liquid phase of the one-band Hubbard model  \cite{Hubbard1963}; its equilibrium properties have been extensively studied using a great variety of different methods  \cite{Metzner1989, Metzner1989B, Georges1993, Montorsi, NATO_HM, Noack2003}. It represents a lattice model which we discuss in the thermodynamic limit, i.e. for an infinite number of lattice sites $N$. We assume translation invariance. Firstly, we do not specify a particular lattice geometry,  work with a generic dispersion relation $\epsilon(k)$ but assume that no nesting of the Fermi surface occurs. Each lattice site is capable of carrying up to two spin - 1/2  fermions. Due to the Pauli principle this implies a local state space of dimension four\footnote{Note the following configurations of the four-dimensional local state space: zero occupation, the spin up or spin-down configuration of single occupancy and the  combination of two antiparallel spins (double occupation).}. Without interaction, all local states are degenerate in energy; hence their composition to the state space of a many-site lattice leads to a single energy band. We discuss the model for finite bandwidth $D$ and in the regime of half filling, i.e. with --on average-- one fermion per lattice site. 
The Hubbard Hamiltonian 
\begin{equation}
\label{INTRO_HM_Ham}
H = -\sum_{ij\sigma} t_{ij} \ c^{\dagger}_{i\sigma}c_{j\sigma} + U \sum_j n_{j \up}n_{j\down} 
-\mu \sum_j \left( n_{j\up}+n_{j\down} \right)
\end{equation}
displays two competing physical processes on the lattice: The first term describes coherent\footnote{An intuitive but not fully correct imagination depicts these hopping processes as an exchange of fermions between different lattice sites. Yet the coherent nature of the hopping is better captured by their delocalizing effect on a single fermionic wavefunction which spreads over more than a single lattice site.} hopping processes between two neighboring lattice sites $j$ and $i$ with site-independent strength $t_{ij}=t=1$ which we set equal to one for convenience. This defines our energy scale in the sequel. The second term depicts a site-independent repulsive interaction $U$. It approximates the influence of a Coulomb repulsion of electrons dwelling on the same lattice site. Due to the Pauli principle, it is effective only between electrons of different spin and is proportional to the product of their spin densities. It reduces the mobility of fermions on the lattice as, intuitively speaking, 'hopping onto singly occupied sites becomes energetically less favorable'. Hence the Hubbard model depicts the competition between delocalizing and localizing effects in an interacting Fermi gas. At zero temperature a Fermi liquid phase exists for weak interaction strength in all dimensions larger than one. 
The chemical potential $\mu$ is set to half filling. 
In a momentum representation it merely accounts for a global energy offset of the kinetic energy of the fermions. The allowed momenta are restricted due to the limited bandwidth $\mathcal{K}= \lbrace \vec{k} \in \mathbf{R}^d: \epsilon(k) \in [-D,D] \rbrace$.  For simplicity, Arabic numbers are used as generic momentum index labels. Note that the Hubbard interaction, while being local in real space, is non-diagonal in momentum space. 
\begin{equation}
\label{HM_INTRO_HAM_MomSpace}
H = \vspace{-1cm} \sum_{k \in \mathbf{K},\sigma \in \lbrace \up, \down \rbrace } \epsilon_k \ {c^{\dagger}_{k\sigma}c_{k\sigma}} \ +\ U   \sum_{1'12'2 \in \mathbf{K}}  {c^{\dagger}_{1'\up}c_{1\up} c^{\dagger}_{2' \down}c_{2\down}} \ \delta^{k_{1'}+k_{2'}}_{k_1+k_2}
\end{equation}

\subsection{Quench of a Fermi liquid}

We implement a particular interaction quench of a Fermi liquid. It is modeled by substituting $U \rightarrow U(t) = \Theta(t) U$ in (\ref{HM_INTRO_HAM_MomSpace}).  

\subsubsection{Normal ordering and energy considerations}
We decompose the interaction term by applying a normal ordering procedure with respect to the ground state of the Fermi gas $\ket{\Omega_0}=\ket{\rm FG}$, which equals the noninteracting Fermi gas (FG). Its momentum distribution is given by $n_k := \bra{\Omega_0} c_k^{\dagger}c_k \ket{\Omega_0}$ and does not depend on spin.  Normal ordered operators are denoted between colons.  
\begin{eqnarray*}
c^{\dagger}_{1'\up}c_{1\up} c^{\dagger}_{2' \down}c_{2\down}
&=& 
\normord{c^{\dagger}_{1'\up}c_{1\up} c^{\dagger}_{2' \down}c_{2\down}} 
+ n_{2'} \delta^{2'}_{2} \normord{c^{\dagger}_{1'\up}c_{1\up}} + 
   n_{1'} \delta^{1'}_{1} \normord{c^{\dagger}_{2'\down}c_{2\down}} 
-  n_{1'} n_{2'} \delta^{1'}_{1} \delta^{2'}_{2} 
\end{eqnarray*}
Thus one particle properties hidden in two-particle scattering terms are extracted and a clear separation of one- and two-particle features in the Hamiltonian is achieved  \cite{Wegner2006}. The one-particle scattering contributions correspond to a shift of the chemical potential of $-U/2$ and could be reallocated to the kinetic energy. However, this additional energy is not dynamically relevant. We stress the point that the observed dynamics is solely caused by two-particle interaction effects. 

Then the sudden switch-on of a normal ordered two-particle interaction term does not lead to a change of the total energy of the system. Yet it \emph{lowers} the ground state energy of the Hamiltonian\footnote{This consequence of an energy conserving quench may seem unphysical, but it highlights a simple fact: While adding a (Coulomb) charge to formerly uncharged particles leads to an additional one-particle potential energy which is first order in $U$ and can be accounted for by a shift of the chemical potential, the corresponding two-particle repulsion (which causes interparticle correlations and lowers the ground state energy) is a second order effect; the dynamics of the quenched Hubbard model is only driven by the later.} such that, at zero time, the noninteracting ground state of the initial  Fermi gas, $\ket{\Omega_0}$, is promoted to a highly excited state of an interacting model. The corresponding excitation energy is measured with respect to the ground state energy of the Hubbard Hamiltonian in equilibrium. It does not vanish in the thermodynamic limit as a single-particle excitation would do. This is because the correlation-induced reduction of the ground state energy becomes effective at \emph{every} lattice site. A more detailed discussion of the involved energies will be resumed in section \ref{HM_SEC_NER}.

\subsubsection{Observables}
To characterize the dynamics of the quenched Fermi liquid we analyze the evolution of particular quantities, namely the total kinetic energy, the total interaction energy and the momentum distribution function. They are expectation values with respect to the initial state $\ket{\Omega_0}$ of the observables $H_0$, $H_{\rm int}$ and the number operator for a fermionic quantum gas 
\begin{equation*}
\label{Intro_Def_N}
\hat{N}_k = \left \lbrace \begin{array}{rcl} c^{\dagger}_k c_k &:& k> k_F \\
1-c^{\dagger}_k c_k &:&k\leq k_F \end{array} \right. 
\end{equation*}
We work in a Heisenberg picture where the observables carry the time dependence.
Notice that the distribution function exhibits the evolution of the occupation of one-particle momentum modes while the energies are mode-averaged quantities.
As the kinetic energy can be easily calculated from the momentum distribution function; and as the total energy of a closed system is conserved, it is sufficient to explicitly calculate the momentum distribution. 

In equilibrium, the zero temperature momentum distribution of an interacting many-particle system of fermions is characterized by a discontinuity at the Fermi momentum $k_F$. Its size, the so-called quasiparticle residue or quasiparticle weight $0\leq Z\leq 1$, reflects the strength of interaction effects: For an interaction-free Fermi gas it acquires its maximum value one, while it decreases with increasing interaction strength in a Fermi liquid.  For the equilibrium Hubbard model this signature has been studied numerically throughout and beyond\footnote{At a critical interaction strength, marking a quantum phase transition from the Fermi-liquid to a non-metallic (Mott insulator) phase, the quasiparticle residue vanishes.}   
the Fermi liquid phase \cite{Mueller-Hartmann1989, Georges1993}. The behavior of the quasiparticle residue under nonequilibrium is the main focus of this work. It is, clearly, a zero temperature analysis as temperature smears out the discontinuity on its own energy scale.

\section{Perturbative analysis of a quenched Fermi liquid}

Similarly to the perturbative study of the squeezed harmonic oscillator we implement the time evolution of the number operator following the scheme of fig. \ref{Fig_Rechenschema} for a Fermi liquid. Again we aim at a perturbative analysis, expanding all results as a power series in the interaction strength. Contrary to the single mode oscillator, the Fermi gas in the thermodynamic limit represents a many-particle problem with an infinite number of degrees of freedom. This implies that many different energy scales contribute to the Hamiltonian. It is obvious that the quench generates occupation in many different excited eigenstates of the interacting Hamiltonian. Therefore we implement the diagonalizing transformation such that a controlled treatment of different energy scales is possible. This can be achieved by a flow equation transformation following Wegner \cite{Wegner1994}. 
Due to the Pauli principle, a fermionic many-particle problem is characterized by the existence of a filled Fermi sea;  the later restricts the phase space for fermionic scattering processes, in particular at zero temperature. This, effectively, reduces the strength of the two-particle Hubbard interaction and allows for the observation of a transient dynamics of an excited state.

\subsection{Flow equation transformation}
\label{HM_SEC_FEQ_FT}

Since it has been independently introduced by Wegner  \cite{Wegner1994} and Glazek and Wilson  \cite{Glazek1993,Glazek1994} the flow equation method became an established tool in the analysis of equilibrium and nonequilibrium many-body systems and has been applied to a great variety of different systems. An extensive list of model systems and problems which have been tackled by the flow equation method can be found in  \cite{Wegner2006} and a comprehensive textbook review is available  \cite{Kehrein_book}. Quite recently, it has been successfully applied to nonequilibrium problems  \cite{Kehrein2005, Hackl2007, Moeckel2008}. For the convenience of the reader we will give a short introduction here. 

\subsubsection{Continuous sequence of infinitesimal transformations}

In section (\ref{QHO_SEC_Trafo}) a single unitary transformation was defined to diagonalize the Hamiltonian approximately. 
The flow equation method, however, decomposes the diagonalization of a many-body Hamiltonian into a \emph{continuous sequence of infinitesimal unitary transformations}. It aims at an approximate diagonalization of the Hamiltonian in energy space. Applied to the nondiagonal Hamiltonian, it imposes a continuous evolution of Hamiltonian parameters. This evolution is, in rough analogy to Wilson's interpretation of the renormalization group, depicted as a flow of the Hamiltonian parameters. 
The flow is parametrized by a scalar, nonnegative and monotonously growing \emph{flow parameter} $B$.

Since only the initial Hamiltonian and, to a lesser degree, the final, approximately energy-diagonal Hamiltonian are fixed boundary conditions there is a large degree of freedom how the continuous sequence of infinitesimal unitary transformations is actually constructed. It allows for the implementation of other desirable features like energy scale separation such that the flow parameter $B$ can be related to an energy scale $\Lambda_B=1/\sqrt{B}$. But contrary to conventional renormalization schemes which distinguish between absolute energies of high and low energy degrees of freedom the flow equation methods separates the treatment of large and small relative \emph{energy differences} in the Hamiltonian. 
This means that those matrix elements of the Hamiltonian which describe deeply inelastic  scattering processes
are eliminated already at an early stage of the flow. 
Successively, those with lower energy differences are treated while elastic scattering processes ('energy-diagonal ones') remain unchanged. Hence, the flow equation method achieves a stable sequence of perturbation theory. 
As all energy scales of the Hamiltonian are retained this motivates the application of the flow equation method to nonequilibrium phenomena.

\subsubsection{Definition of the infinitesimal transformations}
Wegner showed that energy scale separation can be implemented by defining the \emph{canonical generator} of infinitesimal transformations  \cite{Wegner1994}, representing a differential form of equation (\ref{SHO_eta_can_gen_def}). 
\be
\label{HM_eta_can_gen_allg}
\eta(B) = \commutator{H_0(B)}{H_{\text{int}}(B)} 
\ee
It depends on the flow parameter $B$ and is anti-hermitian. The spit-up between the noninteracting and the interacting part of the Hamiltonian has to be defined throughout the flow. 
For the Hubbard Hamiltonian, this is simply achieved by promoting the parameters in the Hamiltonian to 'flowing', $B$-dependent variables. Hence we start with the following ansatz for the flowing Hubbard Hamiltonian
\begin{equation}
\label{HM_HH_MS}
H(t>0;B_0) =  \sum_{k \in \mathbf{K},\sigma \in \lbrace \up, \down \rbrace } \epsilon_k(B) \normord{c^{\dagger}_{k\sigma}c_{k\sigma}} 
+ \sum_{1'12'2 \in \mathbf{K}} U_{1'2'12}(B) \normord{c^{\dagger}_{1'\up}c_{1\up} c^{\dagger}_{2' \down}c_{2\down}}  \delta^{{1'}+{2'}}_{1+2} \quad
\end{equation}

Since the flow of the Hamiltonian is closely related to the net energy difference of scattering processes, an iterative approach will show that the flowing interaction strength inevitably depends on Hamiltonian energies
$U_{1'12'2}(B) \equiv U(\epsilon_{1'}, \epsilon_{2'}, \epsilon_{1}, \epsilon_2;B) $
with an initial condition $U_{1'2'12}(B_0=0) =  U$.
Inserting (\ref{HM_HH_MS}) into (\ref{HM_eta_can_gen_allg}) makes the canonical generator more explicit. With $\Delta \epsilon_{1'12'2}(B) = \epsilon_{1'}(B) -\epsilon_1(B)+\epsilon_{2'}(B)-\epsilon_{2}(B)$ it reads
\begin{equation}
\label{HM_eta_can_gen_spez}
\eta(B) = \sum_{1'12'2 \in \mathbf{K}} U_{1'12'2}(B) \  \Delta\epsilon_{1'12'2}(B) \ \normord{c^{\dagger}_{1'\up}c_{1\up} c^{\dagger}_{2' \down}c_{2\down}} 
\delta^{{1'}+{2'}}_{1+2}
\end{equation}
Still this is an implicit definition of the generator since the functional form of the flowing interaction strength is not known explicitly. 
Hence an iterative approach to the correct and consistent definition of the generator is necessary. It starts with a first parametrization of the flowing coupling constants in (\ref{HM_eta_can_gen_spez}). In a second step its explicit action onto the Hamiltonian allows to calculate an improved parametrization\footnote{This interplay between the definition of the transformation and its action onto the Hamiltonian is a characteristic trait of the flow equation technique. It comes as a consequence both of the demand that the transformation should diagonalize the Hamiltonian and of the very generic construction of the canonical generator in (\ref{HM_eta_can_gen_allg}). The implicit definition allows for a generator which is intrinsically well-adapted to the particular Hamiltonian. }.

\subsubsection{Differential flow equations}
Let $\U(B,dB) = e^{\eta(B) dB}$ be an infinitesimal unitary transformation promoting the Hamiltonian or any other quantum mechanical observable $\O(B)$ to a new representation $\O(B+dB)$. A leading order expansion of its action in $dB$ (which corresponds to the angle $\varphi$ in (\ref{QHO_SEC_Trafo}))
\begin{equation}
\nonumber
\label{FlowEquO}
\O(B+dB) = \U(B, dB) \O(B) \U^{-1}(B,dB) 
\approx \O(B) + \commutator{\eta(B)}{\O(B)} dB
\end{equation}
gives rise to a differential flow equation
\be
\label{HM_FEQ_DEF_O}
\frac{d\O(B)}{dB} = \commutator{\eta(B)}{\O(B)} 
\ee
As a differential statement this is always \emph{exact} and corresponds to the generic definition of a 'generator' in the theory of unitary operations. Note that in (\ref{QHO_SEC_Trafo}) second order contributions in the angle $\varphi$ have been considered which have no equivalence in a differential approach. Approximations enter via the iterative interplay between the generator, the Hamiltonian and other observables. The flow equation method is exact as long as they are considered as abstract, implicitly defined objects. Represented in a  truncated multi-particle operator basis the commutator on the right hand side of equation (\ref{HM_FEQ_DEF_O}) may generate new terms which have not been part of the original representation. Although the later may be extended by these operator terms, this typically runs into an infinite regression  and requires approximate truncations. 
If such a truncation scheme has been established the differential flow equation for operators can be decomposed into a set of coupled differential equations for flowing scalar parameters. We show this first for the transformation of the Hamiltonian which serves to make the canonical generator explicit. All approximations are made with respect to a perturbative expansion of the initial interaction strength $U\equiv U(B=0)$.

\subsubsection{Transformation of the Hamiltonian and the generator in leading order}
We start with the Hamiltonian (\ref{HM_HH_MS}) and a straightforwardly parametrized generator $\eta^{(1)}(B)=\eta(0) \propto U$ [{\it cf.} (\ref{HM_eta_can_gen_spez})]. The first order contribution to the flow equation for the Hamiltonian, i.e. with $\O(B)=H(B)$, comes from the commutator $[\eta(0),H_0] \propto U$ and reads 
\begin{equation*}
\frac{dU_{1'12'2}(B)}{dB} = -U \ (\Delta \epsilon_{1'12'2})^2 
\end{equation*}
It can be integrated and gives a leading order parametrization for the dependence of the flowing interaction strength on energy and on the flow parameter:
\begin{equation}
U_{1'12'2}(B) \quad = \quad U \ e^{-(\Delta\epsilon_{1'12'2})^2B}
\end{equation}
With this parametrization the first-order approximation of the canonical generator reads 
\begin{equation}
\eta^{(1)}(B) = U \sum_{1'12'2 \in \mathbf{K}}   \Delta\epsilon_{1'12'2} \ e^{-(\Delta\epsilon_{1'12'2})^2B} \ \delta^{k_{1'}+k_{2'}}_{k_1+k_2} \times 
 \normord{c^{\dagger}_{1'\up}c_{1\up} c^{\dagger}_{2' \down}c_{2\down}}
\label{HM_FET_GEN_EXPLICIT}
\end{equation}
Its characteristic feature is the exponential cutoff function. It suppresses inelastic scattering processes which violate energy conservation on an energy scale set by $\Lambda_E \sim B^{-1/2}$.

Higher order contributions to the flow of the Hamiltonian include, for example,  the renormalization of the one-particle energies or second order contributions to the flow of the interaction. We will show later that within the accuracy of a second order calculation of the time dependent number operator a leading order implementation of the diagonalization is sufficient; hence they can be neglected and we write (without any further calculation) the energy-diagonal Hamiltonian approximately as
\begin{multline}
\label{HM_HH_EDiag}
H(t>0;B=\infty) =  \sum_{k \in \mathbf{K},\sigma \in \lbrace \up, \down \rbrace } \epsilon_k \normord{c^{\dagger}_{k\sigma}c_{k\sigma}} \ +
\\
+U \sum_{1'12'2 \in \mathbf{K}} \delta(\Delta\epsilon_{1'12'2}) \delta^{k_{1'}+k_{2'}}_{k_1+k_2} \ \normord{c^{\dagger}_{1'\up}c_{1\up} c^{\dagger}_{2' \down}c_{2\down}}
\end{multline}

\subsubsection{Flow equations for the creation operator} 
\label{HM_SUBSEC_FEQ_C}
In a second step we map the creation operator into the approximate energy eigenbasis of the Hamiltonian by means of the flow equation transformation (\ref{HM_FEQ_DEF_O}, \ref{HM_FET_GEN_EXPLICIT}). We define it as a basis independent observable $\C_{k \up}^{\dagger}$ which has different representations depending on the value of the flow parameter. It symbolizes the creation of a physical fermion. In the initial basis of noninteracting fermions, i.e. for $B=0$, it coincides with the formal creation operator $c_{k \up}$ which we treat as the building block of an invariant many-particle operator basis. Represented in this basis,  $\C_{k \up}^{\dagger}$ acquires a composite multiparticle structure under the flow. New terms emerge on the right hand side of its flow equation (\ref{HM_FEQ_DEF_O}) and mirror the dressing of an original electron by electron-hole excitations due to interaction effects. Respecting momentum and spin conservation, this motivates the ansatz
\begin{subequations}
\begin{align}
\nonumber
\tag{\ref{Ansatz-Ca}}
\C^{\dagger}_{k \up}(B) \ =& \   h_{k \up}(B) \ c^{\dagger}_{k \up} 
\\ +&
\label{Ansatz-Ca}
\ \sum_{1'2'1} M_{1'2'1 \up \down \down}(B) \ \delta^{k+1}_{1'+2'} \ 
\normord{c^{\dagger}_{1'\up} c^{\dagger}_{2' \down} c_{1 \down}} 
\\ + &
\label{Ansatz-Cb}
\ \sum_{1'2'1} M_{1'2'1 \up \up \up}(B) \ \delta^{k'+1}_{1'+2'} \ 
\normord{c^{\dagger}_{1'\up} c^{\dagger}_{2' \up} c_{1 \up}}
\end{align}
\end{subequations}
Here $h(B)$ and $M(B)$ are flowing parameters of the observable $\C^{\dagger}_{k \up}(B)$. The first can be linked to the quasiparticle residue via $h_{k_F}(B) = \sqrt{Z(B)}$ and depicts the coherent overlap of the physical fermion with the related interaction-free momentum mode (quasiparticle) of the current representation; the later represents the incoherent background in the spectral function of an interacting system. 
Inserting this ansatz into (\ref{HM_FEQ_DEF_O}) we calculate the differential flow equations  for the flowing parameters $h(B)$ and $M(B)$. A consistent normal ordering of all newly generated terms is essential and causes the emergence of characteristic fermionic phase space factors like 
$Q_{122'}= n_1 n_2 (1-n_{2'}) + (1-n_1)(1-n_2) n_{2'}$.
\begin{eqnarray}
 \pd {h_{k \uparrow} (B)}{B}  
  & = &
  U \ \sum_{1{2^{\prime}}2}\  \Delta\epsilon_{k12'2} \ 
  e^{-B ( \Delta\epsilon_{k12'2} )^2} 
  \label{HM_C_DGL-h} \hspace{0.1cm}
  Q_{1 2{2^{\prime}} } \  
  M_{1 2 {2^{\prime} } \uparrow \downarrow \downarrow}(B)
  \\ 
   \label{HM_C_DGL-Muuu}
  \pd {M_{{{5^{\prime}}}{{6^{\prime}}}{{5}} \uparrow \uparrow \uparrow} (B)}{B} 
  &=&
  - U \sum_{{2^{\prime}} 2} \ [n(2^{\prime}) -n(2)] \  \Delta\epsilon_{2'25'5} \ 
  e^{-B ( \Delta\epsilon_{2'25'5} )^2}
  M_{{{6^{\prime}}}2{2^{\prime}} \uparrow \downarrow \downarrow}(B)
  \\
   \label{HM_C_DEG-Mudd}
  \pd { M_{{{{5}^{\prime}}}{{{6}^{\prime}}}{{{5}}} \uparrow \downarrow \downarrow} (B)} {B} 
   & = & 
   U \sum_{1} 
   \ h_{{1} \uparrow} (B)\ 
   \Delta\epsilon_{5'56'1} \ 
  e^{-B ( \Delta\epsilon_{5'56'1} )^2}  
  \\ \nonumber
   &+&  
  U \sum_{1{2^{\prime}}}  \ \left[ n(1) -n(2') \right]  \
  M_{1 {{6}^{\prime}} {2^{\prime}} \uparrow \downarrow \downarrow }(B) \
  \Delta\epsilon_{2'15'5} \ 
  e^{-B ( \Delta\epsilon_{2'15'5} )^2}
  \\ \nonumber
 &+&  
  U \sum_{12}   \left[  1+ n(2)-n(1) \right]  \ 
  M_{ 1 2 {{5}} \uparrow \downarrow \downarrow} (B)\
  \Delta\epsilon_{5'16'2} \ 
  e^{-B ( \Delta\epsilon_{5'16'2} )^2}
 \\ &+& \nonumber
 U \sum_{1^{\prime}1}  \left[ M_{1  {\tilde{5}} {1^{\prime}} \uparrow \uparrow \uparrow }- M_{{\tilde{5}^{\prime}} 1 {1^{\prime}} \uparrow \uparrow \uparrow } \right]  \left[ n(1^{\prime})- n(1) \right] \ 
 \Delta\epsilon_{1'16'5} \  e^{-B ( \Delta\epsilon_{1'16'5} )^2}
\end{eqnarray}
A perturbative expansion in $U$ of the flowing parameters $h(B)$ and $M(B)$ allows to reduce the complexity of the differential equations but depends on their initial conditions. Since the differential equations are linear, a solution for general initial conditions can be achieved as a linear superposition of solutions for independent initial configurations. We will discuss two cases:
\begin{itemize}
\item [  (A) \ ] \emph{Fully coherent initialization} of one fermion in the momentum mode $k$:  \\
$h_{i\up}(B_0)=\delta^{k}_i$ and $M(B_0)=0$ for all possible indices
\item [ (B) \ ] \emph{Fully incoherent initialization} in the dressing state $p'q'p \up \down \down$: 
$h_i(B_0) = 0$, $M_{1'2'1 \up \down \down}(B_0)=   \delta_{1'}^{p'} \delta_{2'}^{q'} \delta_{1}^{p} $ and $M_{\up \up \up} (B_0)=0$ for all possible indices\footnote{To avoid confusion we stress that in the full context of our problem, this initial condition comes with a pre-factor proportional to $U$. Perturbative arguments always include it. }
\end{itemize}

\subsubsection{Case A: Perturbative analysis at the onset of the flow}
\label{HM_C_A_PA}
We discuss iteratively the action of the differential flow equations at the onset of the flow. In a first step, the flowing parameters $h$ and $M$ on the right hand side of the differential equations can be parametrized by their initial conditions (A). Hence only the first term at the right hand side of (\ref{HM_C_DEG-Mudd}) remains influential and is, due to its pre-factor, of order $U$. 
Consequently, $M_{\up \down \down}(B>0)$ is generated in first order of $U$. Re-inserted into  (\ref{HM_C_DGL-h}) it accounts for a second order correction to the flowing parameter $h(B)$. This describes the leading changes to the quasiparticle residue. Re-inserted into (\ref{HM_C_DGL-Muuu}) and (\ref{HM_C_DEG-Mudd}) it unfolds second order effects on flowing parameters of the $M$ type. Since those do not influence second order results on the momentum distribution function (which will be shown later), this calculation can be, fortunately, dropped. Moreover, the initial conditions (A) are the natural ones to study the behavior of physical fermions. As they do not generate $M_{\up \up \up}$ in relevant order, the  ansatz for $\C^{\dagger}_{k\up}$ can be restricted to 
(\ref{Ansatz-Ca}). This is a consequence of the Pauli principle which disadvantages dressing of a fermion with excitations in the same spin state.  Simplified flow equations read 
\begin{subequations} \label{HM_C_APPROX_DGL_ALL}
\begin{equation}
 \pd {h_{k \uparrow} (B)}{B}  
   = 
  U \ \sum_{1{2^{\prime}}2}\  \Delta\epsilon_{k12'2} \ 
  e^{-B ( \Delta\epsilon_{k12'2} )^2} \ 
  Q_{1 2{2^{\prime}}} \  M_{1 2 {2^{\prime} } \uparrow \downarrow \downarrow}
  \label{HM_C_APPROX_DGL-h} 
\end{equation}
\begin{equation}
\label{HM_C_APPROX_DEG-Mudd}
\pd { M_{{{{5}^{\prime}}}{{{6}^{\prime}}}{{{5}}} \uparrow \downarrow \downarrow} (B)} {B} 
=  
U \sum_{1} 
\ h_{{1} \uparrow} \ 
\Delta\epsilon_{5'56'1} \ 
e^{-B ( \Delta\epsilon_{5'56'1} )^2}  
\end{equation}
\end{subequations}
We stress that this analysis is based on an approximate parametrization of the flowing parameters. It requires that their magnitude under the flow is still sufficiently described by the magnitude of their initial conditions. The following study will show that for the initial conditions of a physical electron at the Fermi energy this remains the case throughout the flow and that, then, non-perturbative effects do not arise.

\subsubsection{Approximate analytical solution of the flow equations}
\label{HM_SEC_C_APPROX_SOL_FEQ}
We integrate the simplified flow equations (\ref{HM_C_APPROX_DGL_ALL}) with respect to the flow parameter $B$. Similar to the above analysis we iteratively develop the flowing parameters from their initial conditions (A). We apply the same parametrization $h_{1 \up} = \delta^{k}_1$ and integrate (\ref{HM_C_APPROX_DEG-Mudd}) from $B_0=0$ to $B_f=B$.
\be
\label{HM_C_Flow_M_2O}
M^{(A), FT}_{{{{5}^{\prime}}}{{{6}^{\prime}}}{{{5}}} \uparrow \downarrow \downarrow}(B)
= U \frac {1-e^{-B ( \Delta\epsilon_{5'56'k} )^2} }{\Delta\epsilon_{5'56'k}}
\ee
This serves as an improved parametrization in (\ref{HM_C_APPROX_DGL-h}) and allows to write down a formal second order correction of $h_{k \up}$. 
\be
\label{HM_C_Flow_h_2O}
h_{k \up}^{(A)}(B) = 1 - U^2 \int_{-\infty}^{\infty}  dE \  \rho \  \frac {(1-e^{-B(\epsilon_k-E)^2})^2}{2 \ (\epsilon_k-E)^2} \ \ I_{k}(E)
\ee
with a phase space factor 
\begin{equation}
I_{k}(E) := \sum_{1'2'1} Q_{1'2'1} \delta(E-\epsilon_{1'} -\epsilon_{2'} + \epsilon_1) \delta^{1'+2'}_{1+k} 
\end{equation}

\subsubsection{Evaluation of the phase space factor}
\label{HM_SubSEC_PSF}
For a further understanding of the flow of $h^{(A)}_{k \up}$ our focus is on the energy dependence of the phase space factor $I_k(E)$. Momentum conservation allows to eliminate the momentum index $1=1'+2'-k$; we keep it as a shorthand notation. We evaluate $I_k(E)$ in energy space assuming a constant density of states $\rho$ 
\be
\label{HM_I_energy_ints}
I_k(E)=
\rho^2 \int_{-\infty}^{\infty} d\epsilon_{1'} \int_{-\infty}^{\infty} d\epsilon_{2'}  
Q(\epsilon_{1'}, \epsilon_{2'},\epsilon_1) \delta(E-\epsilon_{1'} -\epsilon_{2'} + \epsilon_1) 
\ee
and observe that, at zero temperature, the phase space factor $Q(\epsilon_1, \epsilon_{2'}, \epsilon_2)$ vanishes for all but two configurations of its energy arguments \cite{Kehrein_book}:
\begin{center}
\begin{tabular}{|c||c|c|c|c||l|}
\hline
Case & $\epsilon_{1'}$ & $\epsilon_{2'}$ & $\epsilon_1$ & $Q
$ & 
$E= \epsilon_{1'}-\epsilon_1+\epsilon_{2'}$ 
\\ \hline
(a) & $> \epsilon_F$ & $> \epsilon_F$ & $< \epsilon_F$ & 1 & 
$E = +\abs{\epsilon_{1'}} + \abs{\epsilon_{2'}}+ \abs{\epsilon_1} >0$ 
\\
(b) & $< \epsilon_F$ & $< \epsilon_F$ & $> \epsilon_F$ & 1 &
$E = -\abs{\epsilon_{1'}} - \abs{\epsilon_{2'}}- \abs{\epsilon_1} <0$ 
 \\ \hline
\end{tabular}
\end{center}
In both cases the limits of the energy integrations in (\ref{HM_I_energy_ints}) can be restricted by an approximate evaluation of the delta function. Its argument can be rewritten by considering the signs of the one-particle energies $\epsilon_i$ which is done in the last column of the above table. From there one reads off that, since $E$ is a sum of either solely positive or solely negative summands, $E$ forms an upper or lower bound on \emph{both} energy integrations individually. Then both cases (a) and (b) lead to the same approximate expression for the phase space factor 
\be
\label{HM_C_PSF_I_APPROX}
I_k(E)= \rho^2 \int_{\epsilon_F}^E d\epsilon_{1'} d\epsilon_{2'} \ 1 
=\rho^2 (E-\epsilon_F)^2
\ee
We finally remark that the phase space factor $Q(\epsilon_{1'}, \epsilon_{2'},\epsilon_1)$ mirrors the particle-hole symmetry of the Hubbard model. It implies that all odd powers of a perturbative expansion vanish  \cite{Noack2003}. We will see in (\ref{HM_SEC_VILST}) that $Q(\epsilon_{1'}, \epsilon_{2'},\epsilon_1)$ is, in particular, responsible for the suppression of secular terms.

\subsubsection[Analysis of the later flow]{Analysis of the later flow of $h_{k \up}(B)$}
The approximation (\ref{HM_C_PSF_I_APPROX}) allows to interpret the physical relevance of the formal result (\ref{HM_C_Flow_h_2O}): At the Fermi surface, i.e. for $k=k_F$, and at zero temperature its quadratic divergence is cancelled exactly  by the phase space factor.  This implies that the corrections to $h^{(A)}_{k_F \up}$ remain small in second order of $U$ throughout the flow. Hence the parametrization $h^{(A)}_{k_F \up}=1$  used in (\ref{HM_C_A_PA}) and (\ref{HM_SEC_C_APPROX_SOL_FEQ}) is justified even beyond the onset of the flow and the approximate analytical solution gives a correct description of the changes to the quasiparticle residue for all values of $B$. 

Away from the Fermi surface, however, there is no such cancellation by a phase space factor. Now the energy denominator indicates nonperturbative effects. Expanding the exponential in (\ref{HM_C_Flow_h_2O}) shows that the numerator smoothes the peaked energy structure on a $B$-dependent scale such that a pole only emerges in the limit of infinite $B$. 
Nonetheless, the contribution of an environment around $\epsilon_{k \neq k_F}$ to the energy integral in (\ref{HM_C_Flow_h_2O}) grows under the flow until it infers a nonperturbative correction to $h^{(A)}_{k \neq k_F \up}$. This implies a full decay of a particle into incoherent excitations. The corresponding scales can be extracted from a first order expansion in $B$\footnote{A similar analysis based on scaling out the $B$-dependence from the energy integral leads to the same scales but requires a discussion of logarithmic and power-law divergences.}. 
Since this is only justified for $B(\epsilon_k-E)^2 < 1$ we continuously restrict the environment  $[\epsilon_k-1/\sqrt{B},\epsilon_k+1/\sqrt{B}]$ around the later pole. With $e_k = E-\epsilon_k$ and (\ref{HM_C_PSF_I_APPROX}), the correction term in (\ref{HM_C_Flow_h_2O}) has the following dominant behavior for large values of $B$
\begin{multline}
\hspace{-0.25cm}
\Delta h_{k \neq k_F \up}^{(A)} (B)
\approx-U^2 B^2 \frac {\rho^3}2 \int_{\frac {-1}{\sqrt{B}}}^{\frac 1{\sqrt{B}}} d e_k \left(e_k+\epsilon_k-\epsilon_F\right)^2 e_k^2  =
\\ =
-U^2 \rho^3 \left[ \frac 25 \frac 1{\sqrt{B}} + \frac 23 \sqrt{B} (\epsilon_k-\epsilon_F)^2 \right]
\label{HM_C_h_Zerfall_approx}
\end{multline}
The first term can be neglected for large $B$ while the the second term exhibits the decay scale
\be
B^* = \frac 1{U^4 \rho^6 (\epsilon_k - \epsilon_F)^4}
 \quad \text {or} \quad
E^*=U^2 \rho^3 (\epsilon_k -\epsilon_F)^2
\ee
The quadratic energy dependence of $E^*$ resembles the characteristic finite lifetime of Landau quasiparticles away from the Fermi surface $\tau \sim (\epsilon_k -\epsilon_F)^{-2}$. Here it illustrates the complete decay of a physical fermion. This implies that on a scale set by $B^*$ the parametrization of $h_{1 \up} = \delta_1^k$ in (\ref{HM_C_A_PA}) and (\ref{HM_SEC_C_APPROX_SOL_FEQ}) becomes unjustified and the analytical solution (\ref{HM_C_Flow_h_2O}) breaks down. Instead, a better solution of the differential flow equations would show a complete transfer of the spectral weight from the one-particle to many-particle representations.  

This allows to discuss the reliability of our solution: For particles in a perturbatively small environment around the Fermi energy, i.e. $\epsilon_k - \epsilon_{k_F} \approx U$, the energy scale $E^*\sim \Or(U^4)$ is beyond the resolution of a second order calculation. Hence the particle decay cannot be observed on all accessible times. Therefore the extension of the approximate solution (\ref{HM_SEC_C_APPROX_SOL_FEQ})  to an environment around the Fermi energy of radius $U$ leads to a consistent second order result.
Equilibrium calculations have shown that the broadening of the momentum distribution due to interaction effects is concentrated in this region  \cite{Georges1993, Fulde_book}. 

Outside of this environment, particles decay completely into incoherent multiparticle excitations. This implies that, strictly speaking, the validity of (\ref{HM_C_Flow_M_2O}) breaks down. Nonetheless, it shows that under the flow the spectral width $\Gamma_M(B)$ of newly arising incoherent $M$-terms continuously decreases. This indicates that at the decay scale spectral weight is transferred from coherent particles (with a sharp spectral distribution) primarily to excitations with $\Gamma_M \sim E^*$. Since even far away from the Fermi energy this is perturbatively small in $U^2$, the effects of this small widening will not influence the shape of the momentum distribution function. Therefore the solution (\ref{HM_SEC_C_APPROX_SOL_FEQ}) allows for a calculation of the momentum distribution function on all energies.

\subsubsection{Composition of the number operator}
\label{HM_SEC_MDF_NO_composite}
Since the transformation for the creation operator of an electron around the Fermi energy has been established in second order perturbation theory, the number operator $\mathcal{N}_k = \C_k^{\dagger} \C_k$ can be easily composed from the ansatz (\ref{Ansatz-Ca}). The momentum distribution is given by its expectation value with respect to the equilibrium ground state or the nonequilibrium initial state. 
Then only quadratic terms in $h(B)$ and $M(B)$ contribute\footnote{This simple representation of $N_k(B,t)$ depends in a subtle way on the chosen normal ordering prescription. Here the reference state of the normal ordering procedure is identical with that state for which the expectation value of the number operator is taken. If both states disagree, cross-terms proportional to $\sim h \times M$ appear.}:
\begin{multline}
\label{HM_DEF_MDF}
N_k(B,t) = \bra{FG}\mathcal{N}_k(B,t) \ket{FG} =
\\ =
\abs{h_k(B,t)}^2 n_k +
 \sum_{1'2'1} n_{1'} n_{2'} (1-n_1) 
\abs{M^k_{1'2'1 \up \down \down}(B,t)}^2 \delta^{{1'}+{2'}}_{1+k}
\end{multline}
A possible time dependence has already been included for later reference.

\subsection{Equilibrium momentum distribution}
Similarly to the approach in  \cite{Uhrig2002} we observe that the flow equation transformation resembles in many aspects a unitary implementation of Landau's  theory of a Fermi liquid. 
Although a strict identification of the diagonal degrees of freedom obtained from the flow equation approach with Landau quasiparticles is not possible, they motivate an analogous picture of flow equation quasiparticles; like their Landau counterparts they are 
stable only at the Fermi energy and subject to a residual quasiparticle interaction which is carried by the nonvanishing two-particle component of the energy-diagonal Hamiltonian.  Nonetheless, these quasiparticles absorb most of the interaction effects into their definition such that the quasiparticle representation of the interacting ground state is, in good approximation, the filled Fermi sea. 

Hence the momentum distribution in equilibrium can be calculated in a similar way as the equilibrium occupation of the squeezed oscillator (\ref{QHO_NPT_N_EQU}) using (\ref{HM_DEF_MDF}), (\ref{HM_C_Flow_M_2O}) and (\ref{HM_C_Flow_h_2O}).
\begin{eqnarray}
\nonumber
N_k^{EQU} &=& \bra{\Omega} \C_k^{\dagger}(B=0) \C_k (B=0)\ket{\Omega}  
= \nonumber
\bra{\Omega_0} \C_k(B\rightarrow \infty)^{\dagger} \C_k(B\rightarrow\infty) \ket{\Omega_0}  = 
 \\
 &=& 
 -U^2 \int_{-\infty}^{\infty} dE \frac{J_k(E;n)}{(E-\epsilon_k)^2}
  \label{HM_MDF_EQU}
\end{eqnarray}
where the phase space factor 
\begin{multline}
J_k(E;n) = \sum_{1'2'1} \delta^{1'+2'}_{1+k} \ \delta(\epsilon_{1'}+\epsilon_{2'} - \epsilon_{1} - E) \times \\ \times
 [n_k n_1 (1-n_{1'}) (1-n_{2'}) - (1-n_k) (1-n_2) n_{1'} n_{2'} ] 
\end{multline}
resembles the quasiparticle collision integral of a quantum Boltzmann equation \cite{Rammer1986}. 
We will compare (\ref{HM_MDF_EQU}) with the time dependent momentum distribution after an interaction quench.

\subsection{Nonequilibrium momentum distribution}
\begin{figure}
 \begin{center}
  \includegraphics[height=45mm]{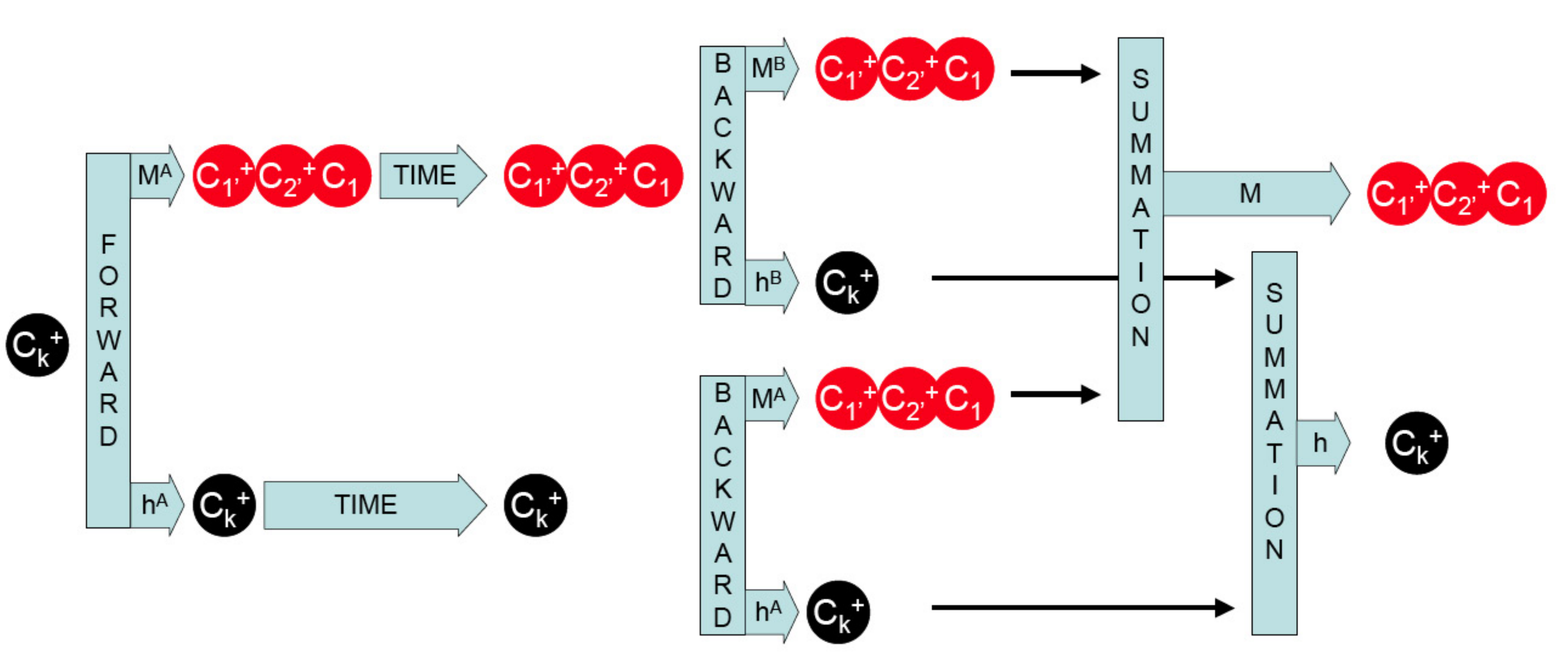} 
    \caption{In nonequilibrium, a fermionic particle $\C^{\dagger}_k$ decays with time. We only consider processes which contribute to the momentum distribution up to second order in $U$. In this schematic diagram the sequence of transformations applied to $\C^{\dagger}_k$ is illustrated. The FORWARD transformation starts with the fully coherent initial conditions of a physical particle. TIME arrows indicate the insertion of phase shifts, the BACKWARD transformation can be decomposed into fully coherent (A) initial conditions and fully incoherent (B) ones. The result of the sequence of transformations is expressed in total coefficients $h$ and $M$.}
  \label{Fig_Zerfall}
  \end{center}
 \end{figure}

We continue our approach towards the nonequilibrium momentum distribution by evaluating the time dependence of the creation operator represented in the initial basis ($B=0$) 
$$
\C_{k \up}^{\dagger}(B=0,t)\ = \   h_{k \up}(0,t) \ c^{\dagger}_{k \up} +
\ \sum_{1'2'1} M_{1'2'1 \up \down \down}(0,t) \ \delta^{k+1}_{1'+2'} \ 
\normord{c^{\dagger}_{1'\up} c^{\dagger}_{2' \down} c_{1 \down}} 
$$ 
according to fig. \ref{Fig_Rechenschema}. Our aim is to write these time dependent coefficients $h(0,t)$ and $M(0,t)$ in section (\ref{HM_SubSEC_CompTR}) as functions of analogously defined (but time independent) parameters of the forward transformation (FT; see \ref{HM_SEC_C_APPROX_SOL_FEQ}) and the backward transformation (BT; see \ref{HM_SubSEC_BT}) as well as of time dependent phase factors introduced by time evolution (\ref{HM_SubSEC_TE}). Fig. \ref{Fig_Zerfall} illustrates some of the used notation and the power counting of the perturbative analysis. 

\subsubsection{Time evolution}
\label{HM_SubSEC_TE}
We time evolve the creation operator in its energy-diagonal representation $\C_{k \up}(B\rightarrow \infty)$ with respect to the energy-diagonal Hamiltonian. 
Its leading part, the noninteracting Hamiltonian, generates a time evolution according to $\mathcal{U}(t,t_0=0)= e^{iH_0t}$ which can be treated exactly. Corrections arise due to the energy-diagonal interacting part of the Hamiltonian and cause the appearance of secular terms.  It will be shown in section (\ref{HM_SEC_VILST})  that they can be neglected in a second order calculation. Hence the action of $\mathcal{U}$ onto the ansatz (\ref{Ansatz-Ca})
\begin{eqnarray}
\nonumber
\C^{\dagger}_{k \up}(B,t) \ &=& \   h^{(A), FT}_{k \up}(B) \ \mathcal{U}^{\dagger}(t,0) c^{\dagger}_{k \up} \mathcal{U}(t,0) 
\\ &+&\label{HM_TIME_Unitary_Ansatz}
\ \sum_{1'2'1} M^{(A), FT}_{1'2'1 \up \down \down}(B) \ \delta^{k+1}_{1'+2'} \ 
\mathcal{U}^{\dagger}(t,0) \normord{c^{\dagger}_{1'\up} c^{\dagger}_{2' \down} c_{1 \down}} \mathcal{U}(t,0)
\end{eqnarray}
is for $B \rightarrow \infty$ fully described by additional phases accompanying the parameters $h^{(A), FT}$ and $M^{(A), FT}$. 
\begin{subequations}
\label{HM_C_Time}
\begin{align}
h_{k\up}(B\rightarrow\infty,t) =& \quad e^{i \epsilon_{k} t} \ h^{(A), FT}_{k'\up}(B\rightarrow\infty) \\
M_{1'2'1 \up\down\down}( B\rightarrow\infty,t) =& \quad 
e^{i(\epsilon_{1'} +\epsilon_{2'} -\epsilon_1)t} \ 
M^{(A), FT}_{1'2'1 \up\down\down}(B\rightarrow\infty)
\end{align}
\end{subequations}

\subsubsection{Inverse transformation}
\label{HM_SubSEC_BT}
The final step depicted in fig. \ref{Fig_Rechenschema} is the backward mapping of the time-evolved observables into the original representation of physical fermions. 
It is implemented by the inverse sequence of differential unitary transformations and simply achieved by interchanging the limits of the $B$-integration in the forward transformation or, intuitively, by 'running the transformation backwards'  \cite{Hackl2007}.  Yet the decay of a physical fermion under the forward transformation has generated nonvanishing incoherent contributions in order $U$; therefore different initial conditions for the backward transformation of $\C_{k \up}(B=\infty, t>0)$ apply. According to (\ref{HM_SUBSEC_FEQ_C}) a linear combination of the solutions for (A) and (B) leads to a full solution at arbitrary initial conditions. Case (A) can directly be taken from (\ref{HM_SEC_C_APPROX_SOL_FEQ}) with $M^{(A), BT} = - M^{(A), FT}$ and invariant $h^{(A)}$. The discussion of case (B) is simplified by noting that the pre-factor of its nonvanishing initial condition is proportional to $U$ because of the generation of this term under the forward transformation. We will consider this additional power of $U$ in a perturbative analysis of the relevant contributions but solve the differential equations for the unweighted initial conditions of case (B).

\subsubsection*{Case B: Perturbative analysis and approximate solution}
At the onset of the backward flow, i.e. for large values of the flow parameter ($B_0=\infty$),  we insert the weighted initial conditions of case (B) as a constant parametrization into the right hand side of the flow equations (\ref{HM_C_DGL-h}-\ref{HM_C_DEG-Mudd}). With $M_{\up \down \down} \sim \Or(U)$, $h_{k \up}(B)$ is generated by (\ref{HM_C_DGL-h}) to second order in $U$; hence, within a second order calculation, the parametrization $h_{k \up}=0$ holds throughout the backward transformation. Consequently, corrections to $M_{\up \down \down}$ are second order in $U$, as well is the generation of $M_{\up \up \up}$. Looking at (\ref{HM_CT_Mges}) and back to (\ref{HM_DEF_MDF}) shows that a second order result of the momentum distribution only requires the knowledge of $M$ to first order. This is a priori known by the weighted initial condition. Integrating   
(\ref{HM_C_DGL-h}) gives
\begin{subequations}
\label{HM_AAS-B-ALL}
\begin{align}
\label{HM_AAS-B-h}
h_{k \up}^{(B), BT}(B) = \ &  U \ Q_{p'q'p} \
\frac{e^{-B(\Delta \epsilon_{p'pq'k})^2}}
{\Delta\epsilon_{p'pq'k}}
\\
\label{HM_AAS-B-M}
M^{(B), BT}_{1'2'1 \up\down\down}(B) = \ &\ 
\delta_{1'}^{p'} \delta_{2'}^{q'} \delta_{1}^{p} 
\end{align}
\end{subequations}

\subsubsection{Composite transformation}
\label{HM_SubSEC_CompTR}
We finish the computation of the time dependent creation operator by composing the forward transformation (FT), the approximate time evolution and the backward transformation (BT) and represent the joint result in terms of time dependent parameters $h_{k \up}(B=0,t)$ and $M_{\up \down \down}(B=0, t)$.
Fig. \ref{Fig_Zerfall} gives a pictorial representation of these expressions.
\begin{subequations}
\label{HM_CT_ALL}
\begin{equation}
h_{k \up}(B=0,t) \ = \  
h_{k \up}^{(A), BT} \
 e^{i \epsilon_{k}t}  \
h_{k \up}^{(A), FT} 
\label{HM_CT_h} 
+\  \sum_{p' q' p} \
h_{k' \up}^{(B), BT} \
e^{i(\epsilon_{p'} +\epsilon_{q'} -\epsilon_{p})t}  \
M^{(A), FT}_{p' q' p \up \down \down} 
\end{equation}
\begin{equation}
M_{1'2'1\up\down\down}^{k'}(B= 0,t) \ =\  
M^{(A), BT}_{{1'}{2'}1 \up\down\down} 
\ e^{i \epsilon_{k}t} \ 
h_{k \up}^{(A), FT} \ 
\label{HM_CT_Mges}
+\  \sum_{p' q' p}
M_{p' q' p \up\down\down}^{(B), BT} \
e^{i(\epsilon_{p'} +\epsilon_{q'} -\epsilon_{p})t} 
M^{(A), FT}_{p' q' p \up\down\down}
\end{equation}
\end{subequations}

\subsubsection{Nonequilibrium momentum distribution}
Inserting the results (\ref{HM_C_Flow_M_2O}, \ref{HM_C_Flow_h_2O}, \ref{HM_C_Time} and \ref{HM_AAS-B-ALL}) via (\ref{HM_CT_ALL}) into (\ref{HM_DEF_MDF}) yields the nonequilibrium time-dependent momentum distribution function for the initial state $\ket{\Omega_0}$. 
\begin{equation}
N^{NEQ}_k(t) := \bra{\Omega_0}\mathcal{N}_k(B=0,t) \ket{\Omega_0} 
\label{HM_Tdep_MDF}
= n_k - 4  U^2 \int_{-\infty}^{\infty} dE \ \frac{ \sin^2( (E-\epsilon_{k}) t / 2 )}{(E-\epsilon_{k})^2}   J_k(E;n)
\end{equation}
We define the correlation-induced time-dependent correction to the momentum distribution 
\begin{eqnarray}
\Delta N^{\rm NEQ}_{k}(t)&=&
N^{\rm NEQ}_{k}(t) - n_{k} \label{NGG-FDF} 
\nonumber 
\end{eqnarray}
and perform a time average of its long-time limit. Note that equation (\ref{HM_Tdep_MDF}) resembles the structure of Fermi's golden rule. There the same energy kernel consisting of the same sinusoidal time dependence and energy denominator becomes increasingly localized with time such that, in the limit of infinite time, the energy integral collapses into a secular term $\propto t$  \cite{Sakurai}.
Here, however, the energy dependence of the phase space factor compensates for the energy denominator at the Fermi energy. This is a particular feature of a many-body fermionic system at zero temperature. Interchanging time average and energy integration then results in a factor of $1/2$. Around the Fermi energy, the long-time limit of the second order perturbative correction to the correlated momentum distribution is given by
\begin{equation}
\label{HM_DeltaMDF_Factor2}
\overline{\Delta N_k^{\rm NEQ}}:= \lim_{t \rightarrow \infty} \langle\Delta N^{\rm NEQ}_{k}(t) \rangle_t =
 - 2U^2 \int_{-\infty}^{\infty} dE {\frac{J_k(E;n)}{(\epsilon_{k}-E)^2}} \
\stackrel{(\ref{HM_MDF_EQU})}{=} \
\mathbf{2} \ \Delta N_k^{EQU}
\end{equation}
Compared with the equilibrium result one observes the key factor two. It represents a similar mismatch of the quasiparticle residue: Its correlation induced reduction is doubled in nonequilibrium compared to the equilibrium result $1-Z^{\rm NEQ}=2(1-Z^{\rm EQU})$.

\subsubsection{Vanishing influence of leading secular terms}
\label{HM_SEC_VILST}
The interpretation of a perturbatively calculated long-time limit result must be based on an analysis of possible secular corrections. Such terms may arise from a simultaneous expansion in both the interaction and time as it was done na\"{\i}vely  by the zeroth order approximation $\mathcal{U}_{B\rightarrow \infty}(t,t_0=0) = e^{iH(B\rightarrow \infty)t} \approx e^{iH_0t}$ in (\ref{HM_SubSEC_TE}); they may render a calculation unreliable on short time scales. We, firstly, notice that up to second order the decomposition 
$$
e^{iHt} = e^{iH_0t} e^{iH_{\rm int}t}  e^{t^2[H_0,H_{\rm int}]/2}\approx e^{iH_0t} e^{iH_{\rm int}t} $$ 
is exact since in the energy-diagonal representation the generator $\eta(B) = [H_0, H_{\rm int}]$ [{\it cf.} (\ref{HM_eta_can_gen_spez})] vanishes.  

Secondly, we will show that the first order expansion of $\Delta \mathcal{U}(t,t_0=0) = e^{iH_{\rm int}(B\rightarrow \infty) t}$, acting onto the ansatz (\ref{Ansatz-Ca}) like $\mathcal{U}$ in (\ref{HM_TIME_Unitary_Ansatz}), does not influence the momentum distribution. 
The further analysis is simplified by the observation that these correction terms (which, of course, solely root in the time evolution of the creation operator) can be \emph{formally} written as time-dependent corrections to the forward transformation. This allows to straightforwardly evaluate their contribution to the full time evolution of the momentum distribution function by inserting the following corrections into (\ref{HM_CT_ALL}) and (\ref{HM_DEF_MDF}):  
\begin{subequations}
\begin{align}
\nonumber
\Delta h^{FT}_{k \up} &= -it \ U \ \sum_{p' q' p} M^{(A), FT}_{p'q'p} Q_{p'pq'}  \delta^{p+k}_{p'+q'} 
\delta(\Delta\epsilon_{p'pq'k})
\\ &
\label{HM_SecularTerms_Dh}
\stackrel{(\ref{HM_C_Flow_M_2O})}{=} i t \  U^2 \int dE \frac{I_k(E)}{E-\epsilon_k} \delta(E-\epsilon_k) \
\stackrel{(\ref{HM_C_PSF_I_APPROX})}{\equiv}  0 
\\
\label{HM_SecularTerms_DM}
\Delta M^{FT}_{1'2'1} &=  -it \ U \ h_{k \up}^{(A)} \delta(\Delta\epsilon_{1'12'k}) \delta^{1+k}_{1'+2'}
\end{align}
\end{subequations}
(\ref{HM_SecularTerms_Dh}) vanishes because of the phase space evaluation presented in (\ref{HM_SubSEC_PSF}). The same argument holds for the influence of (\ref{HM_SecularTerms_DM}) either directly in (\ref{HM_CT_h}) or finally in (\ref{HM_DEF_MDF}). Altogether, there is no leading correction.  

Moreover, 
there is no second order secular term contributing to the momentum distribution. The only one conceivable can be an outcome of the following transformation  
\begin{equation*}
c^{\dagger}_k(B=0) 
\underset{\Or(1)}{\stackrel{\rm FT}{\rightarrow}} c^{\dagger}_k(B=\infty) 
\underset{\Or(Ut)}{\stackrel{H_{\rm int}}{\longrightarrow}} \normord{c^{\dagger}_{1'}c^{\dagger}_{2'}c_1}(B=\infty)  
\underset{\Or(Ut)}{\stackrel{H_{\rm int}}{\longrightarrow}} c^{\dagger}_{k}(B=\infty)
\underset{\Or(1)}{\stackrel{\rm BT}{\rightarrow}} c^{\dagger}_{k}(B=0)
\end{equation*}  
The corresponding correction can be constructed from (\ref{HM_SecularTerms_Dh}) and (\ref{HM_SecularTerms_DM}) and vanishes similar to (\ref{HM_SecularTerms_Dh}). 

Here we observe the suppression of secular terms due to the interplay of time evolution with respect to an energy diagonal Hamiltonian and fermionic phase space factors. It occurs, in most cases, already on the level of the transformation of the creation operator. This illustrates the advantages of the chosen transformation scheme. One secular term, however, only vanishes because of the particular structure of the transformed number operator. 

We conclude that the time evaluation with respect to the noninteracting Hamiltonian $H_0$ in (\ref{HM_SubSEC_TE}) is justified and that the second order long-time limit (\ref{HM_DeltaMDF_Factor2}) is not modified by secular corrections.

\subsection{Nonequilibrium energy relaxation}
Before we discuss the time behavior of the quenched Fermi liquid we establish some results on momentum mode averaged energies like the kinetic energy or the interaction energy. 
\begin{eqnarray}
E^{\text{NEQ,INT}}(t)&:=&U \bra{\Omega_0}H_{\text{int}}(t)\ket{\Omega_0} \\
E^{\text{NEQ,KIN}}(t)&:=&U \bra{\Omega_0}H_0(t)\ket{\Omega_0} = \int_{-\infty}^{\infty} d\epsilon_k \ \epsilon_k \ N_k(t)
\end{eqnarray} 
Although the total energy of a closed system is conserved, its partition onto interaction energy and kinetic energy after the quench is time dependent.

The energy zero point is defined by the Fermi energy of the noninteracting Fermi gas $\epsilon_F= \bra{\Omega_0}H_{0}\ket{\Omega_0} \equiv 0$. It agrees with the kinetic energy shortly after the quench $E^{\text{NEQ,KIN}}(t=0+) = 0$ because the state of the system is not changed by the quench directly and the time evolution of $H_0$ with respect to $H$ has not been effective.
Moreover, as $H_{\text{int}}$ is normal ordered with respect to $\ket{\Omega_0}$, also the interaction energy shortly after the quench vanishes $E^{\text{NEQ,INT}}(t=0+)= 0$.   
This implies that the total energy before and after the quench remains identically zero $E^{\text{NEQ}} = \bra{\Omega_0}H\ket{\Omega_0} := 0$. However, since the ground state of the interacting normal-ordered Hamiltonian is \emph{lower} that that of the Fermi gas, the system is initialized in an excited state.

\subsubsection{Equilibrium energies}
\label{HM_SEC_NER}
 \begin{figure}
 \begin{center}
  \includegraphics[height=45mm]{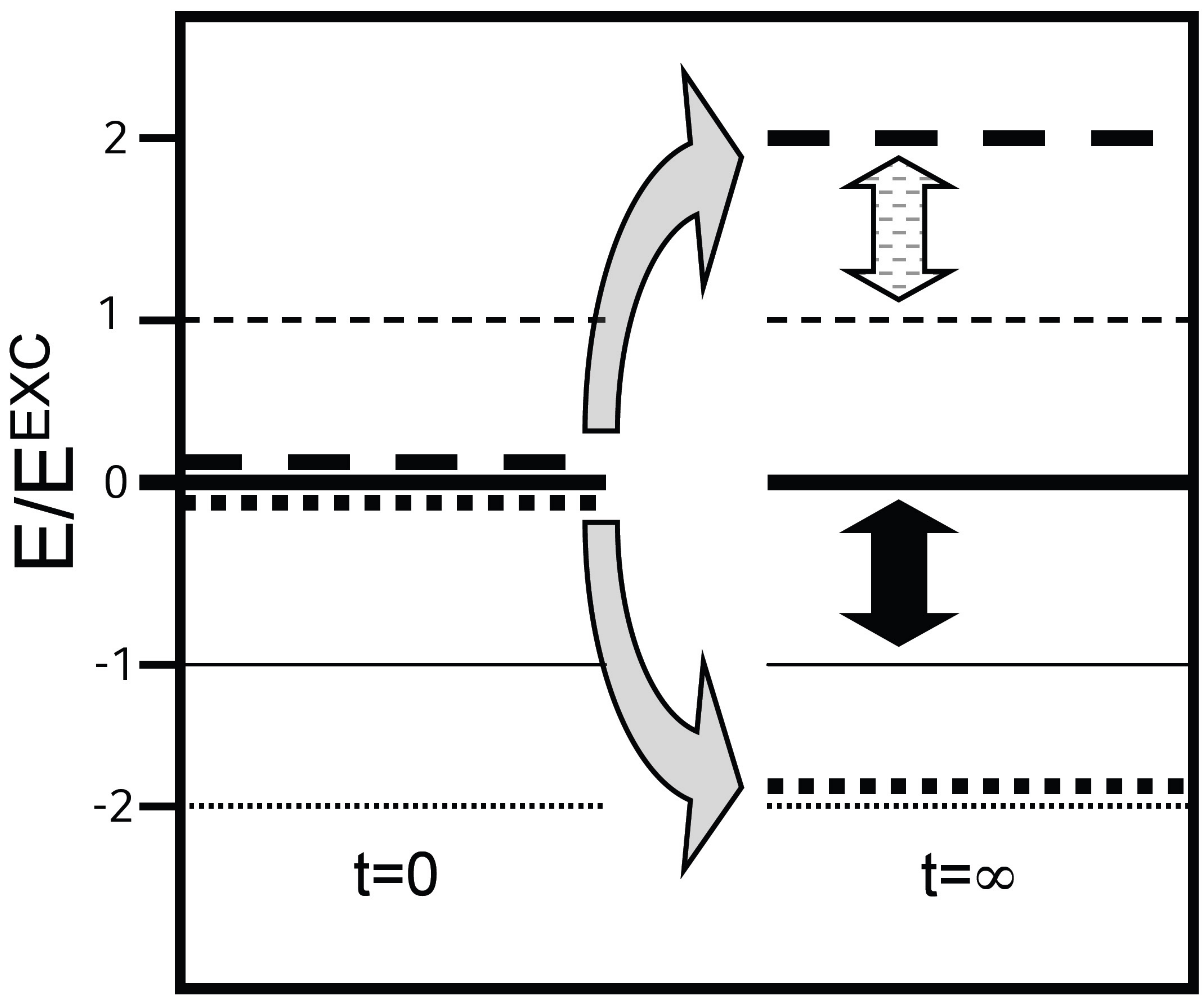} 
    \caption{Sketch of the total energy (full line), kinetic energy (broken line) and interaction energy (dotted line) for the equilibrium (thin lines) and nonequilibrium (thick lines) case in second order perturbation theory. Nonequilibrium energies are compared at the quenching time and in the limit of infinite time. The bowed arrows indicate the corresponding energy relaxation, the straight arrows the gain in kinetic energy (broken filling) which equals the total excitation energy over the equilibrium ground state (full filling).}
  \label{Fig_Energies}
  \end{center}
 \end{figure}
The point of reference of all energy considerations at nonnegative times, however, is the total energy of the equilibrium ground state of the interacting system. We calculate it in perturbation theory with respect to the noninteracting Hamiltonian $H_0$; note that its eigenstates $ \ket{m}$ ($m \in N_0$ with $\ket{\Omega_0}=\ket{0}$) are non-degenerate. 
\begin{equation}
E^{\text{EQU}} := \bra{\Omega}H\ket{\Omega} \stackrel{P.T.}{=} 
\underset{E^{\text{NEQ}}=0}{\underbrace{\bra{0}H_0\ket{0}  + 
\bra{0}H_{\text{int}}\ket{0}}} \ +
\underset{\approx -E^{\text{EXC}}}{\underbrace{\sum_{m \neq 0} \frac{\abs{\bra{0}H_{\text{int}}\ket{m}}^2}{\epsilon_0-\epsilon_m}}} + \Or(U^3)
\label{HM_ENERGIES_EQU_AND_EXC}
\end{equation}
This allows to read off the excitation energy of the quenched quantum system above the interacting ground state $E^{\text{EXC}} := E^{\text{NEQ}} - E^{\text{EQU}} =U^2 \rho_F \alpha \geq 0$.
It is positive, second order in $U$ and its precise value depends -- due to the summation over all quantum numbers $m$ -- on the lattice structure. We hide such details in a numerical prefactor $\alpha$. 

Further equilibrium energies can be calculated from the Feynman-Hellman theorem \cite{CohenTannoudji_book}\footnote{Note that the norm of the interacting ground state is invariant.} 
\begin{eqnarray}
\nonumber
\label{HM_ENERGY_EQU_INT}
\pd{E^{\text{EQU}}(U)}{U} &=& \bra{\Omega(U)}\pd{H(U)}{U}\ket{\Omega(U)} \stackrel{(\ref{HM_ENERGIES_EQU_AND_EXC})}{=} 
 - \frac 2U E^{\text{EXC}} \\
&\stackrel{H}{=} & \bra{\Omega(U)}\frac{H_{\text{int}}}{U}\ket{\Omega(U)} 
= \frac{E^{\text{EQU,INT}}}U
\end{eqnarray}
With (\ref{HM_ENERGIES_EQU_AND_EXC}) and (\ref{HM_ENERGY_EQU_INT}) we know $ E^{\text{EQU,INT}} = -2 E^{\text{EXC}}$ and  $E^{\text{EQU,KIN}}= E^{\text{EXC}}$.  

\subsubsection{Nonequilibrium energies}
Next we calculate the nonequilibrium energies in the long-time limit of the second order calculation. As the Fermi energy is set to zero, 
\be
\overline{E^{\rm NEQ,KIN}} = \int d\epsilon_k \ \epsilon_k \ \overline{\Delta N_k^{\rm NEQ}} 
\stackrel{(\ref{HM_DeltaMDF_Factor2})}= 2 E^{\rm EQU, KIN} = 2 E^{\rm EXC} 
\ee
and $\overline{E^{\text{NEQ,INT}}} = -2 E^{\text{EXC}}$. All energies are sketched in fig. \ref{Fig_Energies}.
Hence a second order calculation shows that the excitation energy of the quench is fully converted into additional kinetic energy such that $\overline{E^{\text{NEQ,KIN}}}=E^{\text{EQU,KIN}}+E^{\text{EXC}}$ while $\overline{E^{\text{NEQ,INT}}}=E^{\text{EQU,INT}}$. This is a remarkable observation.

\subsubsection{Interpretation}
The redistribution of energy between potential and kinetic energy occurs in an early phase of the evolution of the model. Although the full excitation energy is transferred to the kinetic degrees of freedom  this does not mean that the model has thermalized. 
While on average the kinetic energy has acquired its final value the distribution of the energy on the various (momentum) degrees of freedom is still a nonequilibrium one. This can be seen in the momentum distribution function which will relax to thermal equilibrium only on a much longer time scale. In between there is a time regime in which a characteristic nonequilibrium momentum distribution is retained while the energies have already relaxed.


\subsection{Discussion of a multi-step dynamics of the nonequilibrium Hubbard model}

In the following we will illustrate the consequences of the above calculations and develop a picture of the real-time evolution of a Fermi liquid after an interaction quench.
We observe three different time regimes which we illustrate by a numerical evaluation of the time dependent momentum distribution (\ref{HM_Tdep_MDF}). 

\begin{figure}
\centering
  \includegraphics[width=70mm]{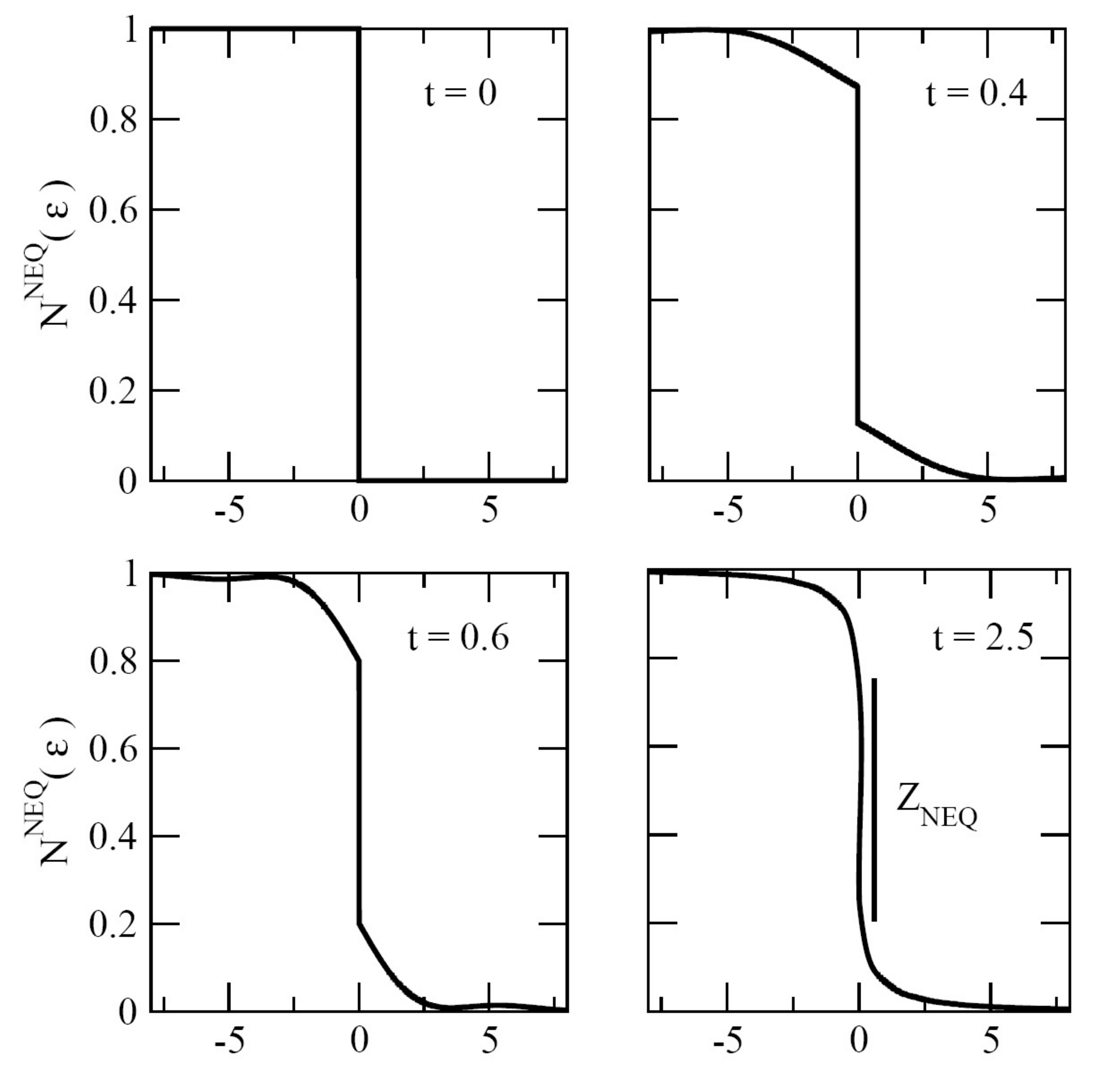}   
  \label{Fig_NGG_FDF}
\caption{The time evolution of the momentum distribution $N^{\rm NEQ}(\epsilon)$ is plotted around the Fermi energy for $\rho_F U=0.6$. A fast reduction of the discontinuity and 1/t-oscillations can be observed. The arrow in the plot for $t=2.5$ indicates the size of the quasiparticle residue in the quasi-steady regime.}
\end{figure}

For computational convenience we use the limit of infinite dimensions where momentum sums can be evaluated by energy integrals. 
It is generally assumed that in this limit the generic features of a Fermi liquid are retained. A Gaussian density of states 
$\rho(\epsilon)=\exp\left(-(\epsilon/t^{*})^{2}/2\right)/\sqrt{2\pi}t^{*}$ includes the constraints of a hypercubic lattice. $t^*$ is linked to the hopping matrix element by dimensional scaling $t \longrightarrow \frac{t^*}{\sqrt{2d}}$ to retain a nontrivial relation between the kinetic and the interaction energy in all dimensions  \cite{Vollhardt1992}. $\rho_{F}=\rho(\epsilon=0)$ denotes the density of states at the Fermi level.
For three time steps explicit results are depicted in fig.~\ref{Fig_NGG_FDF}. 

\begin{figure}
\centering
\includegraphics[width=70mm]{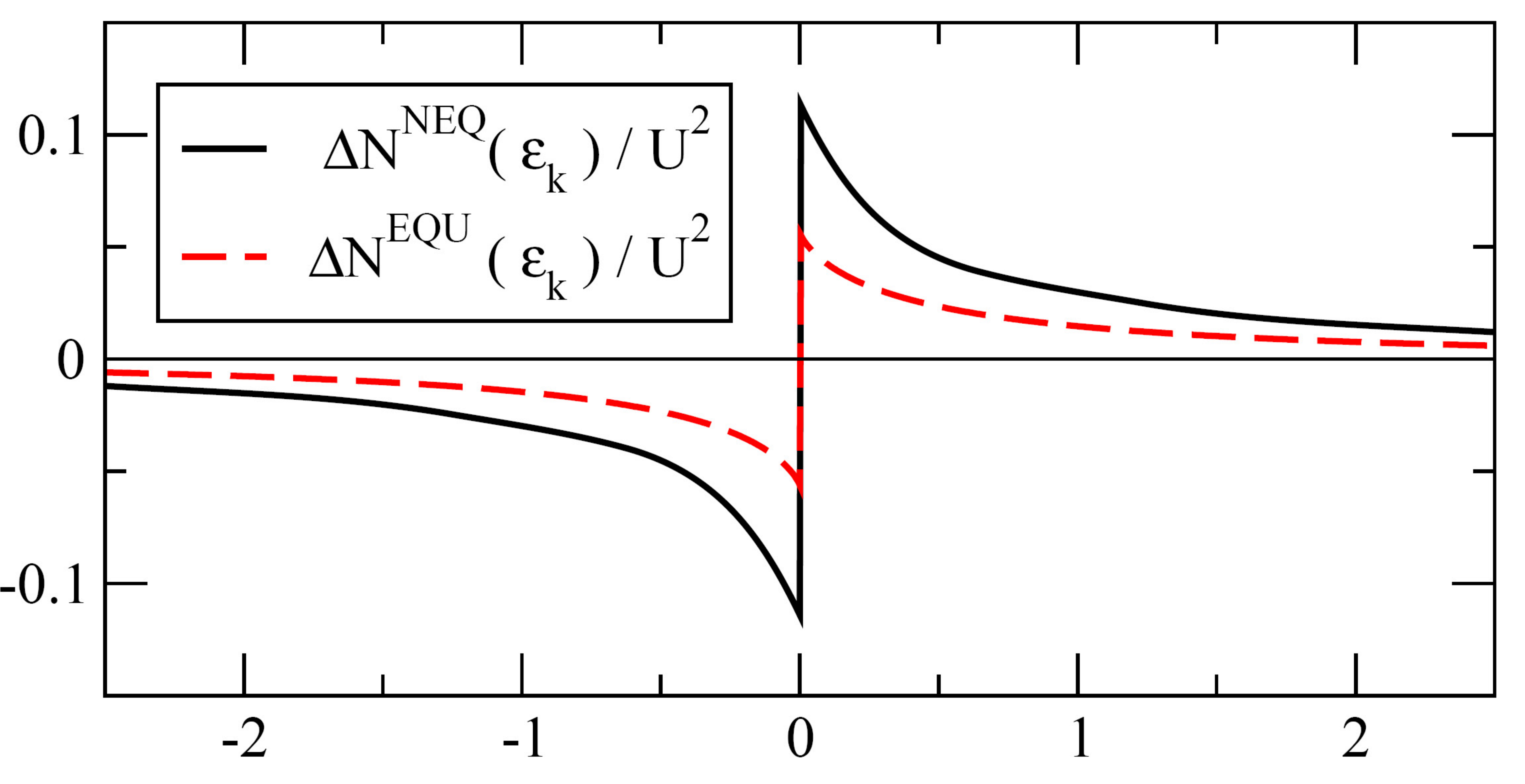} 
\label{Fig_NGG_FDFr}
\caption{The universal curves for the correction to the momentum distribution $\Delta N_{k} = N_k -n_k$ are given for both the equilibrium and for the nonequilibrium quasi-steady state in the weak-coupling limit.}
\end{figure}

\subsubsection{Short-time quasiparticle buildup and nonequilibrium state}
\label{HM_SEC_RESULTS_ST}
The first time regime is the one covered by the above second order perturbative calculation. The evolution of the time dependent \emph{momentum distribution} for physical fermions exhibits the build-up of multi-particle correlations from a noninteracting Fermi gas which leads to the formation of a quasiparticle description. 
The shrinking of the Fermi surface discontinuity of the momentum distribution to a finite, nonzero value (Fig.~\ref{Fig_NGG_FDF} mirrors 
the fast reduction of the quasiparticle residue from one to $Z^{NEQ}<1$ on a time scale is set by second order perturbation theory $0< t\lesssim \rho_F^{-1} U^{-2}$. $1/t$-oscillations accompany this process.

Afterwards, the momentum distribution functions indicates that a zero temperature Fermi liquid description holds. Hence, quasiparticles are well-defined quantities around the Fermi surface. However, the mismatch of the quasiparticle residue by a factor of two (\ref{HM_DeltaMDF_Factor2}) when compared to the corresponding equilibrium distribution displays its continued nonequilibrium nature and a postponed relaxation.  

It is helpful to change the point of view from a momentum distribution of physical fermions (PF) to one of quasiparticles (QP). They agree with each other on the main features. However, since interaction effects are absorbed into the definition of the quasiparticles, the quasiparticle distribution exhibits less pronounced correlation-induced signatures. In equilibrium they vanish completely and a zero temperature quasiparticle distribution always equals a filled Fermi sea with a Fermi surface discontinuity of size one $N^{\rm EQU; QP}_k = \Theta(\epsilon_F-\epsilon_k)$. 

In nonequilibrium, this absorption of interaction effects is incomplete. Due to the increased correlation-dependent reduction of the quasiparticle residue by a factor of two a  mapping of the nonequilibrium momentum distribution for physical fermions into a quasiparticle representation will generate a distribution with a reduced Fermi surface discontinuity. In second order perturbation theory $\overline{N^{\rm NEQ, QP}_k} = N^{\rm EQU, PF}_k$. The reduced discontinuity of the quasiparticle distribution now describes the deviation from equilibrium. We will discuss its further relaxation.  

While, so far, we have only observed a partial relaxation of the momentum distribution, the second order calculation shows a complete transfer of the \emph{excitation energy} from interaction energy to kinetic energy [{\it cf.}~fig.~\ref{Fig_Energies}]. However, since the momentum distribution still exhibits zero-temperature features, this is not related to heating. The simultaneous relaxation of average energies and non-relaxation of other, mode-specific expectation values is known as \emph{prethermalization}. It has been described in nonequilibrium quantum field theories modeling, for instance, the early universe  \cite{Berges2005} and underlines the nonequilibrium nature of the final state depicted by fig.~\ref{Fig_NGG_FDF}.

\subsubsection{Intermediate quasi-steady regime.}

So far we have observed the build-up of a characteristic nonequilibrium state which we have characterized by its energies and momentum distribution.
A study of corrections to the second order perturbative result shows that it extends throughout a second time regime: As the Hubbard model is particle-hole symmetric~ \cite{Noack2003} the next nonvanishing contributions are expected in fourth order. Their relevance for the further dynamics will be discussed later. Here we note that there are no immediate changes to the state of the system for times $t\gtrsim \rho_F^{-1} U^{-2}$. Hence the nonequilibrium Fermi liquid state represents an intermediate quasi-steady regime of the dynamics; for small values of the interaction $U$ it holds for $ \rho_F^{-1} U^{-2}\lesssim t \lesssim \rho_F^{-3} U^{-4}$. This should simplify the observation of this nonequilibrium state in prospective experiments.

\subsubsection{Long-time behavior -- Thermalization}
The mentioned fourth order corrections to the flow equations calculation describe the characteristic dynamics of a third time regime on a scale $ t \sim \rho_F^{-3} U^{-4}$. It originates both from higher order terms of the diagonalizing transformation and from the  full time evolution generated by the energy diagonal interaction Hamiltonian. Their full calculation is beyond the scope of this work. Fortunately, an effective treatment of elastic scattering processes in a quasiparticle representation is provided by the
\emph{quantum Boltzmann equation} (QBE)  \cite{Rammer1986}
\begin{equation}
\pd{N^{\rm QP}_k(t)}{t} = -  \rho_{F}\,U^2\,J_{k}(E=\epsilon_{k},N^{\rm QP}(t)) \ .
\label{HM_RESULTS_QBE}
\end{equation}
Based on its main prerequisite, a well-established quasiparticle picture, it models phenomenologically the relaxation of a nonequilibrium quasiparticle  distribution to a thermal one. One generally expects that it describes the correct long-time behavior although its proper derivation remains an unsolved but well-discussed problem of mathematical physics; this contains the delicate question how the transition from a deterministic quantum dynamics to a irreversible statistical description can be rigorously achieved  \cite{Erdoes2004}. 
Here we only motivate its application.

The characteristic features of the quantum Boltzmann equation can be read off its right hand side which is commonly referred to as the scattering integral \cite{Abrikosov1963}. Since $J_k(E=\epsilon_k, n)$ is energy conserving it describes similar elastic two-particle scattering processes as the energy-diagonal interacting part of the Hamiltonian. Thus a (QBE) description appears as a natural extension of a flow equation analysis which only addresses energy non-diagonal processes. This motivation is backed by the observation of a related coincidence: a perturbative expansion of the scattering integral for the intermediate steady state leads to an analogous fourth order correction as it is expected in the flow equation approach. 
Therefore we link the QBE to the previous dynamics by taking the quasiparticle momentum distribution of the intermediate nonequilibrium Fermi liquid state $N^{\rm NEQ; QP}_{k}$ as its initial condition. 
Because $N^{\rm QP:NEQ}_{k}$ allows nonzero phase space for scattering processes in the vicinity of the Fermi surface, linearizing the phase space factor in the scattering integral shows that the initial quasiparticle distribution function starts to evolve on the time scale $t\propto \rho_F^{-3} U^{-4}$. 
This implies that the quasi-steady fermionic distribution function depicted in the last panel of fig.~\ref{Fig_NGG_FDF}  starts to decay on the same time scale. 

The further dynamics of the quasiparticle momentum distribution function follows, again, from the scattering integral. Since $J_k(E=\epsilon_k, n)$ vanishes for Fermi-Dirac distributions ($n=n^{\rm FD}$) these are the stable fixed points of (\ref{HM_RESULTS_QBE}).  
Moreover, the scattering integral conserves the kinetic energy such that the evolution towards a fixed point is constrained to an energy hypersurface in phase space. 
Hence, if the quantum Boltzmann dynamics continues until it reaches its stable fixed point, thermalization of the momentum distribution can be expected. This implies that the excitation energy, which has relaxed into an excess of kinetic energy already at an earlier stage, is redistributed among the momentum modes until a thermal distribution is  achieved. The corresponding temperature $T_{\rm th} \sim U$ of the thermal momentum distribution follows directly from fitting its Sommerfeld expansion \cite{Ashcroft_Mermin} to the excitation energy.

Notice that fourth order corrections to the diagonalizing transformation may cause an obliteration of the Fermi surface discontinuity even for short times. However, after the quench, the momentum distribution will still show a steep descent; therefore its widening can be safely neglected. We want to mention that the assumption of a quasiparticle picture for all later times is nontrivial. This is a general question of a QBE approach and is usually accepted.

\subsection{Consequences of the Hubbard dynamics}

In the past section we have observed the separation of two time scales of the Hubbard dynamics. While interaction effects lead to a rapid establishment of a quasiparticle picture, the equilibration of the momentum distribution, i.e. heating, is deferred to a much later time. This delayed relaxation is a consequence of two fundamental properties: Firstly, the Pauli principle imposes characteristic phase space restrictions on a multiparticle fermionic system. Those suppress two-particle scattering processes and, thereby, reduce the efficiency of an interaction-driven momentum relaxation; they are responsible for the generic appearance of particle-like low energy excitation physics in a Fermi liquid such that a meaningful analogy between the behavior of the squeezed one-particle oscillator and the many-body Hubbard model can be justified.
Secondly, translational invariance implies the conservation of lattice momenta. Hence there is no other way of momentum relaxation than by momentum transfer in two-particle (or higher) scattering processes. In combination, these two properties form a restrictive bottleneck for the relaxation dynamics. The particular form of the interaction, however, is less important.
For the above calculations we have assumed the applicability of perturbation theory in the interaction strength.
Since only second order and fourth order terms describe the evolution of the momentum distribution, there is no difference between attractive and repulsive interactions; moreover, a generalization to nonlocal interactions is easily possible by introducing momentum dependent interaction matrix elements. The main observation of a characteristic mismatch between the --interaction dependent-- zero temperature correlated equilibrium ground state of the  momentum distribution and a similarly shaped distribution in the intermediate regime of the nonequilibrium case persists.
Hence we expect similar nonequilibrium behavior for a large class of weakly interacting and perturbatively approachable model systems independent of the exact nature and the particular form of the interaction. This reflects the rather generic applicability of Fermi liquid theory for not too strongly interacting systems in equilibrium. 

Moreover, our findings are relevant for studies focussing on the nonequilibrium physics of models with a Fermi liquid instability (FLI). 
Let us consider a quench from a noninteracting Fermi gas into a phase which exhibits such an instability. In the subsequent dynamics on a buildup time $t_{\rm FLI-B}$ one expects both the buildup of the Fermi liquid instability and characteristic nonequilibrium physics related to it. Since non-perturbative weak interaction instabilities are typically linked to exponentially small energy scales, $t_{\rm FLI-B}$ will be large such that characteristic features of the instability are not observable for short times after the quench. In this regime, our perturbative calculation for the nonequilibrium Fermi liquid applies approximately even in the presence of a nonperturbative instability. Quenching into a FLI-phase then requires us to compare the timescales of the dynamics of the nonequilibrium Fermi liquid with that of the instability. 

If the FLI-phase is nonperturbative and distinguished by a gap in the energy spectrum, as it is, for example, the superconducting phase of a Hubbard model with  an attractive interaction, an excitation beyond the energy gap is essential to observe any characteristic nonequilibrium behavior. This excitation can be induced by a sudden interaction quench. As we have seen the inserted energy causes heating effects which may wipe out all signatures of the FLI. 
Still the delayed onset of heating in a Fermi liquid can open a time window for the observation of the nonequilibrium dynamics even in the FLI regime.

A popular example for such behavior is the BCS instability. Recently the study of its  nonequilibrium dynamics following a sudden quench in the BCS interaction has attracted a lot of attention; depending on the precise conditions of the quench, for instance oscillatory behavior in the order parameter $\Delta_{\rm BCS}(t)$ has been found  \cite{Barankov2006, Barankov2006A, Yuzbashyan2006}. These studies only focus on the behavior of the (nonlocal) BCS Hamiltonian which is an effective low energy description of a superfluid. Since its dynamics is integrable a complete topological classification of the behavior of all excited states could be given \cite{Yuzbashyan2006} and no heating is observed.
The actual experimental realization in optical lattices, however, only allows for a quench of the local two-particle Hubbard interaction. Aside from the emergence of an effective BCS interaction, the persistent influence of ordinary Fermi liquid behavior can be expected in such systems. 
Then a quench simultaneously initializes the nonequilibrium dynamics of the instability and heating effects. For a sudden quench heating dominates in agreement with \cite{Barankov2006A} (since $T_{\rm eff} \gg{\Delta_{\rm BCS}}=\exp(-1/\abs{\rho U})$) and makes the nonequilibrium BCS dynamics unobservable. Hence a further analysis of the crossover between instantaneous and adiabatic switching is required to study the visibility of such nonequilibrium dynamics. 

\section{Conclusions}
In this paper we have presented the real-time dynamics following an interaction quench for systems with a discrete energy spectrum and for a Fermi liquid. In all cases we have discussed our key observation which is a discrepancy between the equilibrium and the long-time averaged nonequilibrium occupation by a factor of two. This factor appears precisely in the  modification of the noninteracting occupation due to interaction effects and illustrates a simple example for the interplay of interactions and nonequilibrium conditions. 

In a first analysis this factor was calculated for the squeezed oscillator. Since this is an exactly solvable one-particle model, a comparison between the perturbative and the exact result showed that this discrepancy is not an artifact of perturbation theory. The precise value of two, however, is only reached in the limit of weak interaction. 
Afterwards, this observation was formulated as a theorem applicable to a larger class of discrete systems and observables. We have given two proofs which point out aspects of the origin of the factor two. One of them reconstructs and illustrates the applied transformation scheme which we used to calculate the nonequilibrium dynamics. 

These observations for discrete systems are paralleled by analogous behavior for the zero-temperature momentum distribution function of a quenched Fermi liquid. In the second part of this paper we presented details on a weak interaction quench within the Fermi liquid phase of the Hubbard model in more than one dimension \cite{Moeckel2008}. To cope with the continuous energy spectrum of this Hamiltonian we based our calculation on the flow equation method following Wegner to achieve an approximate diagonalization in energy space. Since the Hubbard model in more than one dimension is nonintegrable, the expectation is that the expectation values of simple observables should thermalize. Hence the characteristic mismatch of the nonequilibrium momentum distribution function when compared with the equilibrium one  
can only be a transient phase of a longer relaxation dynamics. Accordingly, we have found a three step dynamics of the quenched Fermi liquid: During the first phase a nonequilibrium quasiparticle description builds up which is characterized by an untypical value for the nonequilibrium quasiparticle residue. Again, the factor of two appears in the correlation induced reduction of the quasiparticle residue and mimics one-particle behavior. The momentum distribution of this transient nonequilibrium state remains frozen throughout a second phase which can be large for weak interaction. This has been explained by referring to the restricted phase space for two-particle scattering in a zero-temperature fermionic many-body system and corresponds to the collisionless regime. It delays momentum relaxation but does not prevent prethermalization of the kinetic energy. 
A third phase of the dynamics of the momentum distribution follows from the nonequilibrium nature of the transient state. Its nonequilibrium momentum distribution opens phase space for further scattering events. Upholding a quasiparticle picture their backaction onto the momentum distribution is described by a quantum Boltzmann equation which is an effective kinetic equation for the momentum distribution function in terms of a scattering integral. Since the later vanishes for equilibrium distribution functions and nonequilibrium derivations are small in $U^2$ it predicts the thermalization of the momentum distribution on a longer timescale set by $t_{\rm th} \sim \rho^3 U^{-4}$.

This is a major difference to a similar analysis for fermions in one dimension \cite{Cazalilla2006}. There a quench in the forward scattering leads to an integrable dynamics which equals, after bosonization, that one of the squeezed oscillator for each of the momentum modes. In consequence, a similar mismatch between critical exponents of equilibrium and nonequilibrium correlation functions has been found which approaches the factor two in a perturbative limit. However, as no residual quasiparticle interaction occurs, thermalization is impossible: from our point of view our quasi-steady regime extends to all times in this one dimensional model.  Comparing the results for the one dimensional case with our observations for higher dimensions elucidates how thermalization occurs or is inhibited in these two translationally invariant systems.  


Another perspective for further research lies in the examination of the crossover from instantaneous to adiabatic switching. 

\section*{Acknowledgements}
We acknowledge valuable discussions with F. Marquardt. 
This work was supported through SFB 631 of the Deutsche Forschungsgemeinschaft, the Center for NanoScience (CeNS) Munich, and the German Excellence Initiative via the Nanosystems Initiative Munich (NIM). M.M. acknowledges the support of the German National Scholarship Foundation.  

\bibliographystyle{HQL_AoP_1}

\end{document}